\documentclass[final,journal]{IEEEtran}

\usepackage{amsmath}
\usepackage{graphicx}
\usepackage{amssymb}
\DeclareGraphicsExtensions{.eps}
\usepackage{tikz}
\usepackage{xcolor}

\usepackage{pgfplots}
\pgfplotsset{width=5cm,compat=1.9}
\usepgfplotslibrary{external}
\usepgfplotslibrary{statistics}

\usepackage[hyphens]{url}
\usepackage[hidelinks]{hyperref}
\hypersetup{colorlinks=true,linkcolor=black,citecolor=black,filecolor=black,urlcolor=black,breaklinks=true}
\usepackage{cite}
\makeatletter
\def\bstctlcite{\@ifnextchar[{\@bstctlcite}{\@bstctlcite[@auxout]}}
\def\@bstctlcite[#1]#2{\@bsphack
 \@for\@citeb:=#2\do{%
   \edef\@citeb{\expandafter\@firstofone\@citeb}%
   \if@filesw\immediate\write\csname #1\endcsname{\string\citation{\@citeb}}\fi}%
 \@esphack}
\makeatother

\begin{document}
\bstctlcite{IEEEexample:BSTcontrol}
\title{Frequency Response from Aggregated V2G Chargers With Uncertain EV Connections}

\author{Cormac~O'Malley,~\IEEEmembership{Student Member,~IEEE,}%
\ Luis Badesa,~\IEEEmembership{Member,~IEEE,}%
\ Fei Teng,~\IEEEmembership{Senior Member,~IEEE,} \\ %
and Goran Strbac,~\IEEEmembership{Member,~IEEE}%
\vspace{-1.5em}

\thanks{The authors are with the Department of Electrical and Electronic Engineering, Imperial College London, SW7 2AZ London, U.K. (email: c.omalley19@imperial.ac.uk).}
}

\markboth{IEEE Transactions on Power Systems, August~2022}
{Shell \MakeLowercase{\textit{et al.}}: Bare Demo of IEEEtran.cls for IEEE Journals}


\maketitle

\begin{abstract}
Fast frequency response (FR) is highly effective at securing frequency dynamics after a generator outage in low inertia systems. Electric vehicles (EVs) equipped with vehicle to grid (V2G) chargers could offer an abundant source of FR in future. However, the uncertainty associated with V2G aggregation, driven by the uncertain number of connected EVs at the time of an outage, has not been fully understood and prevents its participation in the existing service provision framework. To tackle this limitation,  this paper, for the first time, incorporates such uncertainty into system frequency dynamics, from which probabilistic nadir and steady state frequency requirements are enforced via a derived moment-based distributionally-robust chance constraint. Field data from over 25,000 chargers is analysed to provide realistic parameters and connection forecasts to examine the value of FR from V2G chargers in annual operation of the GB 2030 system. The case study demonstrates that uncertainty of EV connections can be effectively managed through the proposed scheduling framework, which results in annual savings of £6,300 or 37.4~tCO$_2$ per charger. The sensitivity of this value to renewable capacity and FR delays is explored, with V2G capacity shown to be a third as valuable as the same grid battery capacity.
\end{abstract}


\vspace*{-1mm}
\section*{Nomenclature}
\subsection*{Indices and Sets}
\begin{itemize}[\settowidth{\labelsep}{HH} \settowidth{\labelwidth}{HELL}]
\item[$g,G$] Index, Set of generators.
\item[$i,I$] Index, Set of aggregated EV fleets.
\item[$n,N$] Index, Set of nodes in the scenario tree.
\item[$s,S$] Index, Set of storage units.

\end{itemize}

\vspace*{-3mm}
\subsection*{Constants}
\begin{itemize}[\settowidth{\labelsep}{HH} \settowidth{\labelwidth}{HELL}]
\item[$\Delta f_{max}$] Maximum admissible frequency deviation (Hz).
\item[$\Delta \tau(n)$] Time-step corresponding to node $n$ (h).
\item[$\epsilon$] Acceptable risk of FR under-delivery from V2G.
\item[$\eta$] V2G charger (dis)charge efficiency.
\item[$\mu_i$] Mean net EV connection forecast for fleet $i$.
\item[$\pi (n)$] Probability of reaching node $n$. 
\item[$\sigma_i$] Std of net EV connection forecast for fleet $i$.
\item[$c_{LS}$] Value of load-shed from lack of reserve (£/MWh).
\item[$f_0$] Nominal grid frequency (Hz).
\item[$H_g$] Inertia constant of generator type $g$ (s).
\item[$N_{0,i}$] Current number of connected EVs in fleet $i$.
\item[$\Delta \hat{N}_{t}$]  EV connections between $t_d$ and the start of $t_s$.
\item[$P^d(n)$] Total demand at node $n$ (GW).
\item[$P^w(n)$] Total wind power availability at node $n$ (GW).
\item[$P^{sol}(n)$] Total solar power availability at node $n$ (GW).
\item[$RoCoF_{max}$\hspace{-0.5cm}] \hspace{0.5cm} Maximum admissible RoCoF (Hz/s).
\item [$t_d$] Time of scheduling decision (h).
\item [$t_s$] Scheduling time period (h).
\item[$T_{del}$] Delay of FR from EVs (s).
\item [$T_1$] Delivery speed of fast FR (s). 
\item [$T_2$] Delivery speed of slow FR (s).
\end{itemize}

\subsection*{Decision Variables (continuous unless stated)}
\begin{itemize}[\settowidth{\labelsep}{HH} \settowidth{\labelwidth}{HELL}]
\item[$\boldsymbol{b}$] Binary variable to relax (\ref{binvar}).
\item[$\boldsymbol{E_t}$] Aggregate fleet state of charge at time $t$ (GWh).
\item[$\boldsymbol{H}$] System inertia after the loss of $\boldsymbol{PL_{max}}$ (GWs).
\item[$\boldsymbol{N_g}$] Number of committed plants of type $g$ post loss of $\boldsymbol{PL_{max}}$.
\item[$\boldsymbol{P_g}(n)$] Power output of units $g$ at node $n$ (GW).
\item[$\boldsymbol{P^{LS}}(n)$] Load-shed from lack of reserve at node $n$ (GW).
\item[$\boldsymbol{P_s}(n)$] Power output from storage $s$ at node $n$ (GW).
\item[$\boldsymbol{P^{EV}_i}(n)$] Power output from EV fleet $i$ at node $n$ (GW).
\item[$\boldsymbol{P^{wc}}(n)$] Wind curtailment at node $n$ (GW).
\item[$\boldsymbol{P^{solc}}(n)$] Solar curtailment at node $n$ (GW).
\item[$\boldsymbol{PL_{max}}$] Largest power infeed (GW).
\item[$\boldsymbol{R^{ND}}$] Magnitude of fast FR from non-distributed sources (GW).
\item[$\boldsymbol{R^{G}}$] Magnitude of slow FR from thermal plants (GW).
\item[$\boldsymbol{\bar{R}^{EV}}$] Magnitude of scheduled fast FR from all system V2G chargers (GW).

\end{itemize}

\subsection*{Linear Expressions of Decision Variables (Deterministic)}
\begin{itemize}[\settowidth{\labelsep}{HH} \settowidth{\labelwidth}{HELL}]
\item[$\boldsymbol{C_g}(n)$] Operating cost of units $g$ at node $n$ (£).
\item[$\boldsymbol{g_i}$] Individual EV FR capacity in fleet $i$ (GW). 
\item[$\boldsymbol{R^G}(t)$] FR dynamics of thermal plants (GW).
\item[$\boldsymbol{R^{ND}}(t)$] FR dynamics from non-distributed sources (GW).
\item[$\boldsymbol{x},\boldsymbol{u_i},\boldsymbol{v},\boldsymbol{v'}$ \hspace{-0.5cm}] \hspace{0.5cm} Auxiliary expressions for (\ref{Nadir Constraint}) and and (\ref{final form SOC}).
\end{itemize}

\vspace*{-1mm}
\subsection*{Linear Expressions of Decision Variables (Stochastic)}
\begin{itemize}[\settowidth{\labelsep}{HH} \settowidth{\labelwidth}{HELL}]
\item[$\boldsymbol{\delta}$] Excess scheduled FR from EVs compared to amount truly available (GW).
\item[$\boldsymbol{R^{EV}}(t)$] FR dynamics of aggregated V2G chargers (GW).
\item[$\boldsymbol{R^{EV}}$] Magnitude of total available FR from V2G (GW).
\item[$\boldsymbol{R^{EV}_i}$] Magnitude of available FR from fleet $i$ (GW).
\item[$\boldsymbol{y}, \boldsymbol{z}$] Auxiliary expressions for (\ref{Nadir Constraint}).
\end{itemize}

\vspace*{-1mm}
\subsection*{Nonlinear Expressions of Decision Variables (Stochastic)}
\begin{itemize}[\settowidth{\labelsep}{HH} \settowidth{\labelwidth}{HELL}]
\item[$\boldsymbol{\Delta f}(t)$] Frequency deviation at time $t$ after outage (Hz).
\item[$\boldsymbol{t^*}$] Time after outage of frequency nadir (s).
\end{itemize}

\vspace*{-1mm}
\subsection*{Random Variables}
\begin{itemize}[\settowidth{\labelsep}{HH}
\settowidth{\labelwidth}{HELL}]
\item[$\Delta N_i$] Net EV connections for fleet $i$ between $t_d$ and the time of outage during $t_s$.
\end{itemize}

\section{Introduction} \label{Intro}
\IEEEPARstart{D}{ecarbonised} future power systems will be characterised by low inertia due to the displacement of synchronous fossil fuel generators by converter interfaced generation like wind and solar. This makes grid frequency more volatile, thus more challenging to contain within predefined limits. 

Post generator outage, frequency response (FR) is activated to provide a net power injection that aims to arrest frequency decline by restoring the power balance. In this paper, the term FR exclusively refers to this primary FR service within the seconds after a generator loss. The required amount of FR depends on the level of system inertia \cite{Teng2016}, which covers the transient power deficit by extracting kinetic energy stored within the rotating masses of synchronous generators. Previous work \cite{Badesa2019} has demonstrated that fast FR (delivery less than 1s) from converter interfaced resources, like grid batteries, is extremely effective at containing frequency nadir, and thus a vital resource to decouple frequency security from synchronous machines.

Large numbers of electric vehicles (EVs) will be present in future systems, with more than 23 million expected on the road in the UK by 2030 \cite{ClimateChangeCommittee2020}. When paired with vehicle to grid (V2G) chargers, their smart control offers an abundant and valuable \cite{OMalley2020} source of FR. The small size and distributed nature of V2G chargers necessitate they be aggregated together into fleets. The capacity of FR from each fleet is determined by the number of connected EVs, which cannot be known exactly ahead of time. This means that unlike other inverter-based resources like grid-batteries, a fleet's FR capacity is intrinsically uncertain, thus its inclusion makes the dynamic frequency evolution post-outage uncertain.


Most of the current literature focuses on the optimal scheduling of secondary and tertiary frequency response services from EVs. These are steady state services with slower delivery times that are unaffected by low inertia levels and thus transient dynamics do not need to be considered. However, some deterministic methods for scheduling FR do already exist \cite{OMalley2020}\cite{Thingvad2019}\cite{Gao2021}\cite{BLATIAK2022100738}. Reference \cite{Thingvad2019} calculates the profit of FR from individual V2G-connected EVs by calculating the optimal charging schedule in relation to historical FR (delivery within 10s) prices and real EV connection data in Great Britain (GB). The method in \cite{Gao2021} co-optimises an aggregators participation in the day-ahead energy market and the FR market. The only frameworks that unlock the maximum value of FR from EVs by considering fast FR (delivery within 1s) are presented in \cite{OMalley2020} and \cite{BLATIAK2022100738}. Reference \cite{BLATIAK2022100738} optimises a commercial fleet's trip times to maximise FR revenue, whilst \cite{OMalley2020} optimises fleet operation to increase renewable integration in low inertia systems. The paper demonstrates that FR from bidirectional chargers is up to 20 times more valuable than unidirectional chargers.

All these methods assume that the number of EVs that are plugged in at the time of the outage is known. This number determines the power injection capacity of the aggregated fleets and in reality is impossible to know ahead of time. Thus treating it deterministically jeopardises system dynamic security, risking system damage and load shed. To date, the literature that accounts for EV uncertainty \cite{Hajebrahimi2020}\cite{Lu2020}\cite{Amini2020} focuses on hourly energy requirements, which can only be used to manage slow (and less valuable), steady state ancillary services like reserve or dynamic frequency regulation.


This paper offers significant improvements on the current state of the art scheduling methods by explicitly incorporating the intra-hour EV connection uncertainties (which we characterise from field charging data) into the system frequency dynamics. Two stochastic methods are presented that allow, for the first time, the scheduling of FR under this uncertainty. This contribution unlocks the substantial value of FR from EVs in future systems whilst maintaining guarantees on system dynamic security.

A simple `individual' approach is presented that limits the scheduled FR from each fleet individually, similar to the approach that the UK's system operator takes for aggregated FR providers \cite{NationalGridESO2022}. The main contribution of this paper is a second `joint' approach that schedules an aggregate amount of FR across all the diverse EV fleets on the system. The second approach focuses on ensuring system dynamic security and offers significant improvements over the `individual' approach that we demonstrate mathematically and with case studies.

The `joint' approach we propose schedules FR from fleets of EVs with uncertain plugins using distributionally robust chance constraints (DR-CC). These allow low-probability violation of uncertain constraints for a set of possible probability distributions called an ambiguity set.
DR-CC finds the balance between stochastic and robust approaches. It leverages distributional information like moment or unimodality knowledge, to result in less conservative results than robust programs, but requires less precise distributional knowledge than stochastic programs. Furthermore, many useful ambiguity sets facilitate highly tractable analytical convex reformulations.

Ambiguity set construction is generally categorised into two distinct approaches, moment based \cite{Lu2020}\cite{ Amini2020}\cite{Zhang2017}\cite{Bagchi2021}\cite{Roald2015}\cite{Xie2018} and statistical distance based \cite{Chen2018,Zhou2020}, with some recent work seeking to combine the two \cite{Hajebrahimi2020}. Both approaches have seen widespread application within steady state energy system modelling, primarily to deal with renewable power generation forecast uncertainty \cite{Hajebrahimi2020}\cite{ Amini2020}\cite{Bagchi2021}\cite{Roald2015}\cite{Xie2018} \cite{Chen2018} \cite{Zhou2020}.

Recently, moment based DR-CC are also increasingly being employed to deal with the intrinsic uncertainties of aggregated distributed resources (ADRs) \cite{Zhang2017, Bagchi2021, Amini2020 ,Lu2020}.
Reference \cite{Lu2020} develops a method for distribution systems to mitigate their renewable power forecast uncertainty via aggregated EV charging. A model predictive control scheduling approach is used, with uncertain EV charging demands accounted for via moment based DR-CC. DR-CCs are used in \cite{Amini2020} to facilitate the provision of fast reserves from aggregated behind-the-meter loads (including EVs and water heaters). Uncertain energy and power constraints on reserve are considered, with the option to exploit distributional unimodality information to tighten the ambiguity set. Reference \cite{Zhang2017} applies DR-CCs to schedule reserve from aggregated air-conditioning loads with uncertain reserve capacity limits within an optimal power flow problem. Ambiguity sets considering exact and approximate second moment information are used, which result in a Second-Order Cone (SOC) Program and a Semi-Definite Program respectively. Finally, Bachi \textit{et al.}~\cite{Bagchi2021} apply the conic reformulation of a two-sided linear DR-CC with known second order moments, first derived in \cite{Xie2018}, to line loading and nodal voltage constraints under uncertain renewable outputs and uncertain ADR energy demand, revealing the impact of network constraints on ADR's bidding strategy in the day ahead electricity markets. 

However, the above references  only utilise the flexible demand of ADRs (such as EVs) to provide reserve for steady state power balancing. Non consider the ability of ADRs to assist in the dynamic problem of securing frequency in the transient period immediately following the loss of a large generator. 
In this paper we establish a DR-CC method to optimally schedule FR from aggregated EV fleets. The operator can specify frequency security violation probability, allowing the preferred trade-off between system risk and operational cost reduction to be found. To the best of our knowledge, this is the first work to explicitly evaluate the impact of uncertainty on the value of FR from distributed providers. Although this paper exclusively focuses on EVs, the presented method is also applicable to other ADRs.


The main contributions of this work are:
 \begin{enumerate}
     \item To investigate the impact of EV connection uncertainty on their provision of primary (fast) FR in the transient period immediately following an outage. The uncertainty is incorporated into the system frequency dynamics, from which probabilistic nadir and steady state security frequency requirements are derived. 
      \item To propose a novel, convex moment-based DR-CC on the maximum scheduled FR from V2G chargers.  This convexifies the probabilistic frequency security constraints whilst enabling the valuable scheduling of FR from V2G-connected EVs in an efficient and risk-limited manner. 
     \item To analyse field EV fleet connectivity data to provide realistic parameters and EV connection forecasts, as well as guide ambiguity set selection.
     \item To provide new insight, based on simulating the yearly operation of the GB 2030 system, on the the value of FR from V2G and its sensitivity against renewable generation penetration, grid battery penetration, communication delays and uncertainty levels.  
 \end{enumerate}

This paper is organised as follows: Section~\ref{Agg EV Fleet Model} derives a convex formulation for probabilistic frequency security constraints. 
Field EV fleet connectivity data is analysed in Section~\ref{EV Connectivity Forecasting} to inform EV connectivity forecasting and parameter selection, whilst Section~\ref{Case Studies} presents case studies exploring the value of FR from aggregated V2G chargers. Section~\ref{conclusions} gives the conclusions.


\section{Modelling of Aggregated V2G Chargers in Frequency Dynamics} \label{Agg EV Fleet Model}
This section derives frequency security constraints from the dynamic swing equation, which are non-deterministic due to the uncertain FR capacity from V2G chargers. The proposed DR-CC formulation for these constraints is presented, along with the virtual battery model of aggregate fleet charging.
\vspace*{-3mm}
\subsection{Frequency Security Constraints Under Uncertainty}\label{frequency security cons}
System frequency evolution post generator loss is accurately approximated by the single machine swing equation \cite{KundurBook}: 
\begin{equation}
\label{swng equation}
\frac{2 \boldsymbol{H}}{f_0} \frac{d \boldsymbol{\Delta f}}{d t} = \boldsymbol{R^{EV}}(t) + \boldsymbol{R^{ND}}(t) + \boldsymbol{R^G}(t) - \boldsymbol{PL_{max}}
\end{equation}
Load damping is neglected as the level in future systems dominated by power-electronics will be much reduced \cite{Chavez2014}. Thermal plants are grouped by generator types. The system inertia is determined by the number of committed thermal plants: 
\begin{equation}
\label{inertia definition}
\boldsymbol{H} = \sum_{g \in G} H_g \cdotp P^{max}_g \cdotp \boldsymbol{N_g}
\end{equation}
The formulation is compatible with binary commitment variables for each individual generator. However, previous work \cite{Sturt2012} has shown that due to the large number of generators considered, grouping the binary commitment variables within each generator group into one continuous commitment variable $\boldsymbol{N_g}$ significantly improves problem solve times with a negligible impact on results \cite{Sturt2012}. This approach is consistent with the literature \cite{Teng2016,Badesa2019,OMalley2020}. 

FR dynamics are modelled as linear ramps, similar to the work in \cite{Chavez2014,Teng2016,Badesa2019, Badesa2020}. Detailed dynamic simulations carried out in Section III of \cite{Badesa2020} show that droop controls can be accurately and conservatively approximated by a ramp. More detailed dynamic models prohibit closed form solutions to (\ref{swng equation}), necessary in order to derive convex algebraic frequency security constraints. 
\begin{equation}
\label{Unit ramp}
\boldsymbol{R^{EV}}\hspace{-1mm}(t) = 
    \begin{cases} 
      \frac{\boldsymbol{R^{EV}}}{T_{1}}\cdotp t & t \leq T_{1} \\
      \boldsymbol{R^{EV}} & t > T_{1} \\
    \end{cases}
    ,
    \boldsymbol{R^{ND}}\hspace{-0.5mm}(t) = 
    \begin{cases} 
      \frac{\boldsymbol{R^{ND}}}{T_{1}}\cdotp t & t \leq T_{1} \\
      \boldsymbol{R^{ND}} & t > T_{1} \\
    \end{cases}
\end{equation}
\begin{equation}
\label{Unit ramp 2}
\boldsymbol{R^G}(t) = 
    \begin{cases} 
      \frac{\boldsymbol{R^G}}{T_{2}}\cdotp t & t \leq T_{2} \\
      \boldsymbol{R^G} & t > T_{2} \\
    \end{cases}
\end{equation}    
In this paper $T_1 < T_2$. The slower speed models governor controlled FR from thermal plants. The faster speed comes from power-electronic devices, including V2G chargers and non-distributed devices like grid batteries.

Ahead of time, the dynamics of the cumulative FR delivered from V2G chargers is known. However, the magnitude of delivered FR is uncertain because this is proportional to the number of connected EVs which cannot be known ahead of time. Thus, the grid frequency dynamic is also uncertain. For an individual fleet the response capacity is determined by the charging decisions of that fleet (decision variables), and the number of EVs that are connected (random variable):

\begin{equation} 
    \boldsymbol{R_{i}^{EV}} = \underbrace{(D_{max,i} - \boldsymbol{D_i} + \boldsymbol{C_i})}_{\boldsymbol{g_i}} \cdotp (N_{0,i} +  \Delta N_i )
\end{equation}
The cumulative magnitude of FR from all fleets is:

\begin{equation} \label{R1 total def}
   \boldsymbol{R^{EV}} = \sum_{i\in I} \boldsymbol{R_{i}^{EV}} 
\end{equation}


It is assumed that charging decisions for EVs within the same fleet are uniform. The number of currently connected EVs $N_0$ is known. The net EVs connected between now and the time of generator outage ($\Delta N$) can be forecast, but not known exactly in advance. This paper presents a stochastic framework to incorporate FR from aggregated V2G chargers, whilst explicitly limiting the risk of frequency security breach due to potential under-delivery of FR from EVs.

\subsubsection{RoCoF Constraint}


The maximum RoCoF occurs at the moment of $\boldsymbol{PL_{max}}$ outage. At this time no response has been delivered so it is deterministic and limited by inertia alone. Constraining the maximum RoCoF is necessary to prevent RoCoF-sensitive protection systems from disconnecting distributed generation and exacerbating the deficit. Setting $t=0$ in (\ref{swng equation}) results in:
\begin{equation}
\label{RoCoF.2}
  \frac{2 |RoCoF_{max}|}{f_0} \cdotp \boldsymbol{H} \geq \boldsymbol{PL_{max}}
\end{equation}
\subsubsection{Steady State}
Frequency drop will be arrested if the sum of FR is greater than the largest loss. Ensured to a high certainty with:
\begin{equation}
\label{steady state chance}
    \mathbb{P} \bigg[ \boldsymbol{PL_{max}} \leq \boldsymbol{R^{ND}} + \boldsymbol{R^{EV}}  + \boldsymbol{R^G} \bigg] \geq 1 - \epsilon
\end{equation}
\subsubsection{Nadir Constraint}
Here it is assumed that the nadir occurs after $T_1$ ($\approx 1$s) as the extremely low inertia required to breach the frequency limit $\Delta f_{max}$ ($\approx -0.8$Hz) would violate the RoCoF constraint (\ref{RoCoF.2}) for realistic power system parameters. The frequency nadir occurs at the instant of zero RoCoF. According to (\ref{swng equation}) this is:
\begin{equation}
\label{Nadir time}
\boldsymbol{t^*} = \frac{[\boldsymbol{PL_{max}}-(\boldsymbol{R^{ND}} + \boldsymbol{R^{EV}})]\cdotp T_2}{\boldsymbol{R^G}}
\end{equation}
It is shown in \cite{Badesa2019} that by integrating (\ref{swng equation}) and then substituting in (\ref{Nadir time}), the nadir constraint can be formed as a convex rotated SOC. Thus the post outage frequency drop is contained with high assurance via:
\begin{multline}
\label{Nadir Constraint}
    \mathbb{P} \Bigg[ \bigg(  \underbrace{ \frac{\boldsymbol{H}}{f_0}-\frac{(\boldsymbol{R^{ND}} + \boldsymbol{R^{EV}}) \cdotp T_1}{4 \Delta f_{max}} }_{= \ \boldsymbol{z}} \bigg)\hspace{-0.1cm}\underbrace{ \frac{\boldsymbol{R^G}}{T_2} }_{= \ \boldsymbol{x}} \hspace{-0.1cm} 
    \\ \geq \hspace{-0.1cm} \bigg(\underbrace{\frac{\boldsymbol{PL_{max}} - (\boldsymbol{R^{ND}} + \boldsymbol{R^{EV}}) }{2 \sqrt{\Delta f_{max}}}}_{= \ \boldsymbol{y}}  \bigg)^2 \Bigg] \geq 1 - \epsilon
\end{multline}

Finally, the power injection from aggregated chargers may be delayed due to communication or frequency measurement lag. An additional term in the nadir constraint can account for this \cite{Badesa2020}:
\begin{equation}
\label{Delay definition}
\boldsymbol{z} = \frac{\boldsymbol{H}}{f_0}-\frac{(\boldsymbol{R^{ND}} + \boldsymbol{R^{EV}}) \cdotp T_1 }{4 \Delta f_{max}} - \frac{\boldsymbol{R^{EV}} \cdotp 2 T_{del} }{4 \Delta f_{max}}
\end{equation}




\subsection{Convex Reformulation of Chance Constraints} \label{reformulation section}
The inclusion of response from aggregated EV's within the frequency dynamics make (\ref{steady state chance}) and (\ref{Nadir Constraint}) non deterministic. Therefor they cannot be applied to optimisations within scheduling or market contexts. This severely limits their use and motivates the need for their convex and deterministic reformulation. Making (\ref{Nadir Constraint}) deterministic is challenging as no convex analytical reformulation of a chance constrained SOC currently exists.

Overcoming this mathematical dead-end to produce a convex reformulation of (\ref{Nadir Constraint}) and (\ref{steady state chance}) that maintains the guarantees on frequency security under uncertain FR delivery is the main methodological contribution of this paper. It is achieved via the insight that, because $\boldsymbol{R^{EV}}$ is the only non deterministic parameter within (\ref{steady state chance}) and (\ref{Nadir Constraint}), ensuring that they are met with `$(1-\epsilon)$\%' certainty is equivalent to scheduling an amount of response from EVs ($\boldsymbol{\bar{R}^{EV}}$) that will be delivered with `$(1-\epsilon)$\%' certainty. This is found via:


\begin{equation} \label{R1 uncertainty}
   \mathbb{P} \Bigg[ \boldsymbol{\bar{R}^{EV}}\leq \sum_{i\in I} \boldsymbol{R_{i}^{EV}} \Bigg] \geq 1-\epsilon
\end{equation}
$\boldsymbol{\bar{R}^{EV}}$ replaces $\boldsymbol{R^{EV}}$ in (\ref{steady state chance}) and (\ref{Nadir Constraint}), making them deterministic. The argument within chance constraint (\ref{R1 uncertainty}) is linear, hence if the mean and standard deviation of the forecasted number of connected EVs is known, it can be analytically reformulated into a SOC \cite{Roald2015}. Thus a deterministic convex formulation is achieved.
We start by defining a new scalar random variable: 
\begin{equation} \label{R1 uncertainty 3}
   \boldsymbol{\boldsymbol{\delta}} = \boldsymbol{\bar{R}^{EV}} - \sum_{i\in I}\boldsymbol{g_i}\cdotp(N_{0,i} +  \Delta N_i )
\end{equation}
Note, $\boldsymbol{\boldsymbol{\delta}}$ represents the excess scheduled FR compared to the FR actually available. We want this to be negative with high probability. In a deterministic system $\Delta N_i$ is known, $\boldsymbol{\bar{R}^{EV}} = \sum_{i\in I} \boldsymbol{R_{i}^{EV}}$ and thus $\boldsymbol{\boldsymbol{\delta}} = 0$.

Substituting (\ref{R1 uncertainty 3}) into (\ref{R1 uncertainty}):
\begin{equation} \label{R1 uncertainty 1}
   \mathbb{P} \bigg[ \boldsymbol{\boldsymbol{\delta}} \leq 0 \bigg] \geq 1-\epsilon
\end{equation}
Assuming independent EV connections between fleets, the mean and standard deviation of $\boldsymbol{\boldsymbol{\delta}}$ are:
\begin{equation} \label{R1 uncertainty 4}
   \boldsymbol{\mu}(\boldsymbol{\boldsymbol{\delta}}) =\boldsymbol{\bar{R}^{EV}} - \sum_{i\in I} \boldsymbol{g_i}\cdotp (N_{0,i} + \mu_i) ,\ \boldsymbol{\sigma}(\boldsymbol{\boldsymbol{\delta}}) = \sqrt{\sum_{i \in I} (\boldsymbol{g_i} \cdotp \sigma_i)^2}
\end{equation}
Subsequently, $\boldsymbol{\boldsymbol{\delta}}$ can be scaled to have a zero mean and unit variance via $\boldsymbol{\boldsymbol{\delta}}_n~=~ [\boldsymbol{\boldsymbol{\delta}}-\boldsymbol{\mu}(\boldsymbol{\boldsymbol{\delta}})]/ \boldsymbol{\sigma}(\boldsymbol{\boldsymbol{\delta}}) $:
\begin{equation} \label{R1 uncertainty 2}
   \mathbb{P} \bigg[ \boldsymbol{\boldsymbol{\delta}}_n \leq \frac{-\boldsymbol{\mu}(\boldsymbol{\boldsymbol{\delta}})}{\boldsymbol{\sigma}(\boldsymbol{\boldsymbol{\delta}})} \bigg] \geq 1-\epsilon
\end{equation}
The cumulative distribution function ($F_\mathcal{P}(k)$) of $\boldsymbol{\boldsymbol{\delta}}_n$ gives the probability that $\boldsymbol{\boldsymbol{\delta}}_n$ takes a value less than or equal to some constant $k$:
\begin{equation}
    F_\mathcal{P}(k) = \mathbb{P}[\boldsymbol{\boldsymbol{\delta}}_n \leq k]
\end{equation}
We now consider the DR-CC formulation, an important strength of this method is that the exact and true distribution $\mathcal{P}$ of $\boldsymbol{\boldsymbol{\delta}}_n$ does not need to be known. The set of possible distributions that $\mathcal{P}$ might belong to is called an ambiguity set ($\mathbb{A}$), defined by the distributional assumptions made on $\mathcal{P}$. However, because $\mathcal{P}$ is not known exactly, the exact form of its cumulative distribution is also unknown, inhibiting the reformulation of (\ref{R1 uncertainty 2}). Following the method presented in \cite{Roald2015}, this problem can be overcome by defining a lower bound on $\mathcal{P}$'s cumulative distribution function
($f_\mathcal{P}(k)$): 

\begin{equation} \label{cdf lower bound}
  f_\mathcal{P}(k) = \textup{inf}_{\mathcal{P} \in \mathbb{A}} \ F_\mathcal{P}(k)
\end{equation}
Given that $F_\mathcal{P}(k) \geq f_\mathcal{P}(k) \ \forall \ k$, then its substitution into (\ref{R1 uncertainty 2}) maintains the inequality and guarantees that the scheduled response is deliverable with at least the specified certainty level:
\begin{equation}  \label{cdf equivalence}
    F_\mathcal{P}(k) \geq  f_\mathcal{P}(k) \geq 1-\epsilon
\end{equation}



Given that the function of $f_\mathcal{P}(k)$ is increasing, it has a well defined inverse $f^{-1}_\mathcal{P}(\lambda)$. Thus constraint (\ref{R1 uncertainty 2}) can be written as:
\begin{equation} \label{non expanded full write}
  \frac{-\boldsymbol{\mu(\boldsymbol{\boldsymbol{\delta}})}}{\boldsymbol{\sigma}(\boldsymbol{\boldsymbol{\delta}})} \leq f^{-1}_\mathcal{P} (1-\epsilon)
\end{equation}
Given that the RHS of (\ref{non expanded full write}) is a constant, this constraint is a convex SOC after substituting in the moments of $\boldsymbol{\boldsymbol{\delta}}_n$ (\ref{R1 uncertainty 4}):
\begin{equation}
\label{final form SOC}
    \sqrt{\sum_{i \in I} (\underbrace{\boldsymbol{g_i} \cdotp \sigma_i}_{\boldsymbol{u_i}})^2} \leq \underbrace{ \frac{1}{ f^{-1}_\mathcal{P}(1-\epsilon)} \cdotp \Big(\sum_{i \in I} \boldsymbol{g_i}\cdotp (N_{0,i} + \mu_i) - \boldsymbol{\bar{R}^{EV}} \Big)}_{\boldsymbol{v}}
\end{equation}

For some instances when risk aversion is high and the standard deviations of $\Delta N_i$ are large, constraint (\ref{final form SOC}) can be infeasible. To ensure feasibility during simulation, a binary variable is added to the right hand side. Using the big-M technique, it simultaneously relaxes (\ref{final form SOC}) and constrains $\boldsymbol{\bar{R}^{EV}}$ to zero.
\begin{equation}
\label{binvar}
    \sqrt{\sum_{i \in I} (\boldsymbol{u_i})^2} \leq \underbrace{\boldsymbol{v} + M_1 \cdotp \boldsymbol{b}}_{\boldsymbol{v'}}
\end{equation}

\begin{equation}
    \boldsymbol{\bar{R}^{EV}} \leq M_2 \cdotp (1-\boldsymbol{b})
\end{equation}
The exact form of $f^{-1}_\mathcal{P}(\lambda)$ depends on the distributional assumptions made about $\mathcal{P}$. Here, all ambiguity sets assume knowledge of the first two moments of $\boldsymbol{\boldsymbol{\delta}}$. According to (\ref{R1 uncertainty 4}), this requires the true mean and standard deviation of forecasted EV connection numbers ($\mu_i,\sigma_i$) for each fleet to be known. Three distinct ambiguity sets are considered:
\begin{enumerate}
    \item Distributionally Robust Optimisation (DRO) - When only the mean and standard deviation of $\boldsymbol{\boldsymbol{\delta}}$ are known.
    \item Unimodal - The distribution of $\boldsymbol{\boldsymbol{\delta}}_n$ is assumed to have a single peak. Likely when $\Delta N_i$ are themselves unimodal.
    \item Gaussian -  The distribution of $\boldsymbol{\boldsymbol{\delta}}_n$ is assumed to be Gaussian. True when $\Delta N_i$ are themselves Gaussian.
\end{enumerate}
More statistical information on $\mathcal{P}$ defines tighter ambiguity sets and thus results in a less conservative chance constraint. The distributional assumptions made on $\boldsymbol{\boldsymbol{\delta}}$ define the form of $f^{-1}_\mathcal{P}(1-\epsilon)$, according to the inverse cumulative distribution for the Gaussian case, or probability inequalities for the Unimodal and DRO case. The exact expressions for $f^{-1}_\mathcal{P}(1-\epsilon)$ are shown in Table \ref{f expressions}.

\begin{table}[!t]
\renewcommand{\arraystretch}{1.3}
\caption{Expressions for $f^{-1}_\mathcal{P}(1-\epsilon)$}
\label{f expressions}
\centering
\begin{tabular}{|c||c|}
\hline
Gaussian & $f^{-1}_\mathcal{P}(1-\epsilon) = \Phi^{-1}(1-\epsilon) $\\
\hline
Unimodal & $f^{-1}_\mathcal{P}(1-\epsilon) = 
\begin{cases} 
      \sqrt{\frac{4}{9 \epsilon} - 1} & \textup{for} \ 0 \leq \epsilon \leq \frac{1}{6} \\
      \sqrt{\frac{3(1-\epsilon)}{1+3\epsilon}} & \textup{for} \ \frac{1}{6} \leq \epsilon \leq 1 \\
\end{cases}
 $ \\
\hline
DRO & $f^{-1}_\mathcal{P}(1-\epsilon) = \sqrt{\frac{1-\epsilon}{\epsilon}}$\\
\hline
\end{tabular}
\vspace*{-3mm}
\end{table}

A great strength of our proposed constraint to schedule response from EVs under uncertainty is that it results in a convex programme. Post substitution of $\bar{R}^{EV}$ into (\ref{steady state chance}) and (\ref{Nadir Constraint}), the RoCoF (\ref{RoCoF.2}) and the steady-state (\ref{steady state chance}) constraints are linear. The deterministic nadir constraint (\ref{Nadir Constraint}) is a rotated SOC of the form $\boldsymbol{z} \cdotp \boldsymbol{x} \geq \boldsymbol{y}^2$. Finally the `joint' constraint to limit $\bar{R}^{EV}$ (\ref{final form SOC}) is a standard SOC of the form $||\boldsymbol{u}||_2 \leq \boldsymbol{v'}$ \cite{boyd2004convex}, where $\boldsymbol{u}$ is a vector of $\boldsymbol{u_i}$. Convexity allows it to be applied to a wide range of market and scheduling problems and solved in polynomial time by widely available commercial solvers with guaranteed convergence. It also allows the use of dual variables for shadow pricing.

\subsection{Comparison to Simplistic Risk-Aware Scheduling Method}
This is the first paper to present a framework to schedule fast FR from aggregated EVs under connection uncertainty, so comparison to state of the art is not straightforward. However, some operators already allow the participation of aggregated resources in FR markets \cite{NationalGridESO2022} under strict individual deliverability guarantees. We formulate this mathematically, for the first time, by applying (\ref{R1 uncertainty}) to schedule response from each fleet individually with `$(1-\epsilon)\%$' deliverability probability: 
\begin{equation} \label{Indv 2}
   \mathbb{P} \Bigg[ \boldsymbol{\bar{R}_i^{EV}}\leq \boldsymbol{g_i} \cdotp (N_{0,i} +  \Delta N_i ) \Bigg] \geq 1-\epsilon \hspace{0.5cm}\forall i
\end{equation}
The sum of these equals the cumulative FR from EVs on the system:
\begin{equation} \label{Indv 1}
   \boldsymbol{\bar{R}}^{EV} = \sum_{i\in I} \boldsymbol{\bar{R}_{i}^{EV}} 
\end{equation}
Due to only one uncertainty ($\Delta N_i$) being present in each constraint of (\ref{Indv 2}), according to (\ref{final form SOC}) their reformulation is linear:
\begin{equation}
\label{Indv 3}
    \boldsymbol{g_i} \cdotp \sigma_i \leq  \frac{1}{ f^{-1}_\mathcal{P}(1-\epsilon)} \cdotp \Big( \boldsymbol{g_i}\cdotp (N_{0,i} + \mu_i) - \boldsymbol{\bar{R}^{EV}_i} \Big) \hspace{0.1cm} \quad \forall i
\end{equation}
Application of (\ref{Indv 3}) is referred to as the `individual' method, and represents the current state of the art. 

The primary concern of an operator is to guarantee that the frequency is contained with a high probability. For the `individual' method, the relationship between choice of `$(1~-~\epsilon)~\%$' (e.g.~99\%) for individual fleets and the guarantee on the total system dynamic security is not known. Indeed if they decrease the risk of individual fleet under delivery of FR past the desirable system level, the guarantee on system dynamic security is lost. For this reason when comparing the methods we use the same $\epsilon$ value.

The improvement of our proposed `joint' method over the incumbent `individual' method can be mathematically quantified by subtracting the scheduled FR ($\bar{R}^{EV}$) in (\ref{Indv 1}) from the amount scheduled using (\ref{final form SOC}):
\begin{equation}
\label{Indv 4}
    \Delta \boldsymbol{\bar{R}^{EV}} = f^{-1}_\mathcal{P}(1-\epsilon) \cdotp \Bigg( \sum_{i \in I} \boldsymbol{g_i} \sigma_i - \sqrt{\sum_{i \in I} (\boldsymbol{g_i} \sigma_i)^2} \Bigg)
\end{equation}
For any set of positive real numbers the root of the sum of squares will always be less than or equal to the sum of those numbers. Thus $\Delta \bar{R}^{EV}$ will always be greater than or equal to zero. In real terms this means that our proposed method will always schedule more response from the system's EVs than the current state of the art whilst maintaining dynamic security. 

Conceptually this is because the `joint' method compared to the `individual' method leverages the fact that on the rare occasions ($\approx 1$\%) when one fleet has many fewer EVs plugged in than forecast and under delivers FR, the other fleets are ($\approx99$\%) likely to compensate by over delivering their scheduled FR. Thus the `joint' method's focus on the system dynamic security makes better use of the EV FR resource for more efficient operation.

\subsection{State of Charge and Reserve}\label{SOC Section}
Contrary to when scheduling FR from EVs, scheduling charging uses hourly EV connection numbers. These are assumed deterministic and known. This is justified because: 1) As shown in Section \ref{Case Studies} and \cite{OMalley2020}, an EV's value in highly renewable systems is dominated by its FR provision. Thus, characterising the impact of uncertainty on EV value is unhindered. 2) Despite constraint (\ref{final form SOC}) being fully compatible with charging-under-uncertainty methods (such as that presented in \cite{Amini2020}), deterministic EV charging is used here to simplify modelling and increases insight into system operation under (\ref{final form SOC}), the core contribution of the paper.

Here, each EV fleet is modelled as a virtual aggregate battery. All fleets are modelled in the same way, so the subscript $i$ is dropped for notational clarity. A battery's charge rate is equal to the sum of all the individual constituent EVs:
\begin{equation} \label{EV power equation}
    \boldsymbol{P^{EV}} = (N_{0} + \Delta \hat{N}_{t}) \cdotp (\boldsymbol{D} - \boldsymbol{C})
\end{equation}
$\Delta \hat{N}_{t}$ is the net change in EV numbers between now and the beginning of the scheduling period $t$. It is deterministic and read into the simulation via a time series. EV connections are discretized, occurring at the start of each hour, then remaining constant for the entire timestep. This makes $\Delta \hat{N}_{t}$ different to $\Delta N$ in (\ref{R1 uncertainty 3}), which is a random variable, representing the instantaneous number of EVs connected. This varies within the hour time-step, necessary because response capacity is determined by the instantaneous net power injection capacity of fleets. Whereas a virtual battery's generation and state of charge relate to the averaged connectivity values.

A battery's state of charge at the end of a scheduling period depends on: that period's charge decisions; its parent node's state of charge at the end of the previous timestep ($\boldsymbol{E_{t-1}}$); and change in charge incurred by EV (dis)connection.
\begin{equation} \label{EV charge equation}
    \boldsymbol{E_{t}} = \boldsymbol{E_{t-1}} + (N_{0} + \Delta \hat{N}_{t}) \cdotp (\eta \boldsymbol{C} - \frac{1}{\eta} \boldsymbol{D}) + \Delta N^{in}_t  E^{in} - \Delta N^{o}_t  E^{o}
\end{equation}
Note that $\Delta N^{in}_t, \Delta N^{o}_t$ refer to the number of EVs that connect and disconnect at the beginning of timestep $t$ respectively. Thus $\Delta \hat{N}_{t}$ is their cumulation between now and the scheduling time period $\Delta \hat{N}_{T} = \sum_{t=1}^T (\Delta N^{in}_t - N^{o}_t) $. An EV's (dis)connection states of charge ($E^{o},E^{in}$) are assumed known. Note $E^{o} \geq E^{in}$, thus $E^{o} - E^{in}$ represents the EV driving energy expenditure.
\vspace*{-1mm}
\subsection{Stochastic Unit Commitment} \label{SUC Section}
A pre-existing scheduling model is significantly enhanced to optimally co-ordinate charge and generation decisions in light of uncertain future EV connections. This model is used for simulations of annual system operation under different constraints on the use of FR from V2G connected EVs. These simulations demonstrate: the value of our proposed formulation; insight into its impact on system operation (e.g. change in wind curtailment); and validates the frequency security risk guarantees. This section briefly introduces the advanced stochastic unit commitment (SUC) model.

\begin{figure}[!t]
\vspace*{-1mm}
\centering
\includegraphics[width=0.97\linewidth]{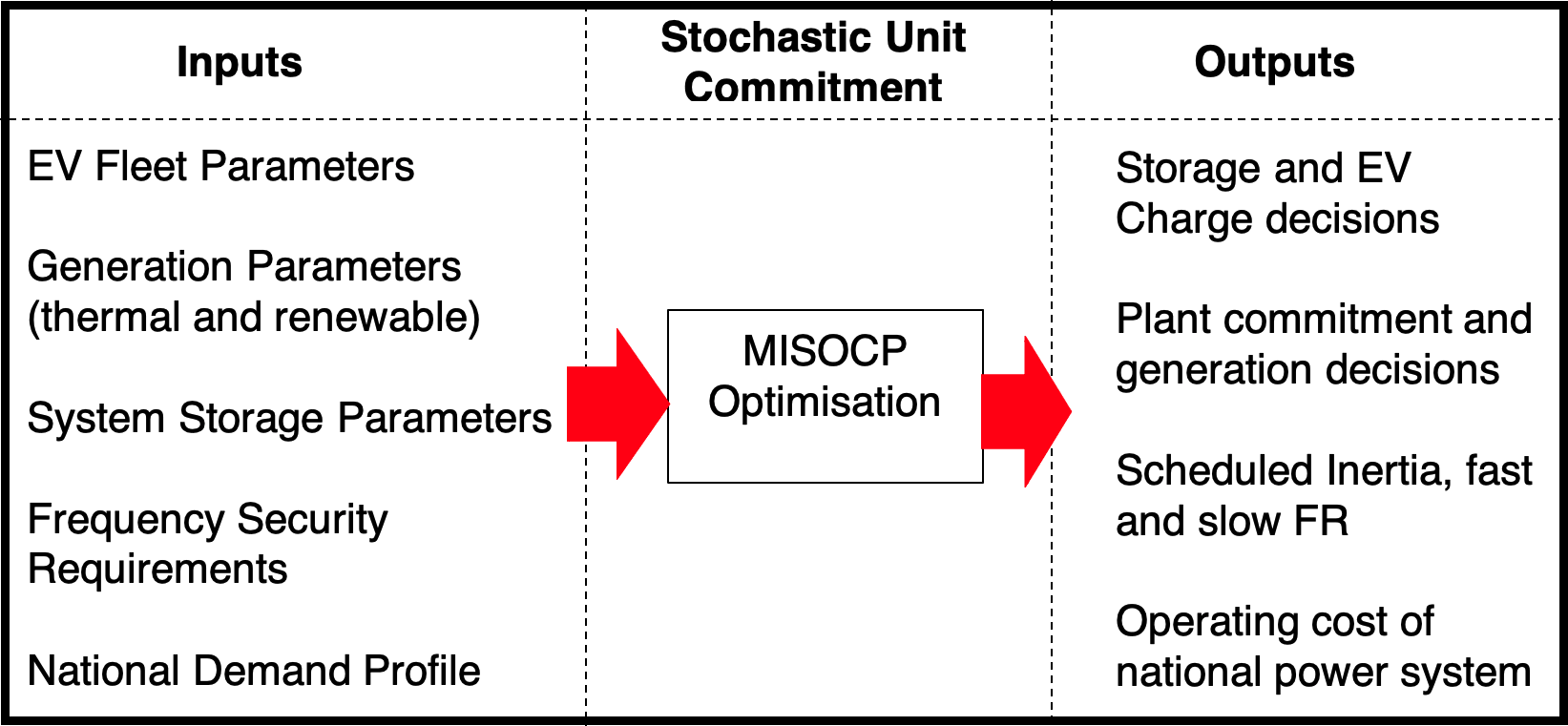}
\vspace*{-2mm}
\caption{Graphic representation of the input/output flow for the SUC. To simulate annual system operation, system variables are updated and the optimisation is iterated every hour.}
\label{SUC picture}
\vspace*{-1mm}
\end{figure}

The SUC model optimally schedules generator and storage actions to provide reserve, response (fast and slow), inertia and energy production under uncertain renewable output over a 24hr period. Fig.~\ref{SUC picture} represents the process graphically.

User defined quantiles of the random variable of net demand (demand net of wind power) are used to construct a scenario tree. Reference \cite{Sturt2012} demonstrates that well chosen quantiles that branch only at the root node can discretize the continuous range of potential wind realisations in an accurate manner whilst yielding a highly tractable model. 

A single-bus power system model is used. The justification for this is two-fold: 1) It is common to solve the commitment and dispatch problems sequentially, adjusting the initial solution to account for line-flows. Here we model the first stage problem. 2) Frequency is a system-wide quantity, so insight into the contribution of FR from V2G connected EVs does not change significantly by including multiple buses.

Each node has a complete set of decision variables, subject to standard generator and storage constraints including minimum stable generation, maximum (dis)charge, state of charge limits and the power balance constraint:
\begin{multline}
\label{power balance}
    \sum_{g \in G} \boldsymbol{P_g}(n) + \sum_{s \in S} \boldsymbol{P_s}(n) + \sum_{i \in I} \boldsymbol{P^{EV}_i}(n) + P^w(n)- \boldsymbol{P^{wc}}(n)
    \\ + P^{sol}(n) - \boldsymbol{P^{solc}}(n) = P^d(n) - \boldsymbol{P^{LS}}(n)
\end{multline}
There are also inter timestep constraints between nodes to define states of charge, plant commitment times and minimum up/down times. An exhaustive constraint list can be found in Section III of \cite{Sturt2012}. All constraints are linear other than the nadir and `joint' (\ref{final form SOC}) constraint to limit $\boldsymbol{\bar{R}^{EV}}$. Thus the SUC is a mixed-integer second order cone programme (MISOCP). 

The probability of reaching a given scenario (node) weights the cost function:
\begin{equation}
\sum_{n\in N} \pi(n) \bigg( \sum_{g\in G} \boldsymbol{C_g}(n) + \Delta \tau (n) (c^{LS} \boldsymbol{P^{LS}}(n)) \bigg)\label{Cost Function}
\end{equation}
A rolling planning approach is used to simulate annual system operation. Decisions that minimise the expected operating cost over the next 24h period are found. The decisions at the current root node are implemented and the system is rolled forwards by an hour, updating system states as well as wind and EV connectivity realisations. With this new information the scenario tree is updated and the process iterated.

\section{EV Connectivity Forecasting and Data Analysis} \label{EV Connectivity Forecasting}
Scheduling FR from aggregated V2G chargers requires forecasting the number of connected EVs, which is equivalent to accurate characterisation of the distributions of $\Delta N_i$. This is important for two main reasons: 1) The analytical reformulations of (\ref{final form SOC}) and (\ref{Indv 3}) require knowledge of the true mean and standard deviation (std); and 2) The $\Delta N$ distributions inform ambiguity set selection. 

Constraint (\ref{final form SOC}) translates the specified risk of under delivery to a scheduled amount of FR. It is compatible with any forecasting technique that ascertains the mean and std of $\Delta N$. Here, a simple forecasting technique to characterise $\Delta N$s using real data is adopted. The future connectivity is assumed well characterised by data from a similar time of the week during the previous year. This is to demonstrate how forecast outputs are translated into operational inputs. It is expected that in actual operation more advanced forecasting techniques will be employed.


\subsection{Forecasting Technique}

Real, open source EV fleet charging data \cite{truedata} for 2017 is used to characterise two distinct fleets, `domestic' and `work'. The domestic fleet relates to 3.2m charging events across 25,000, 10kW chargers installed in people's homes. The work fleet relates to 103,000 charging events across 540, 20kW chargers installed in car parks of public sector buildings. Most of the individual chargers only provided data for specific months of the year. To account for this, the number of active chargers each month was found, `active' defined as at least 2 charge events per week. The true fleet size was then taken as the average number of active monthly chargers, 8,500 and 200 chargers for domestic and work, respectively. 

\begin{figure}[!t]
\centering
\includegraphics[width=0.97\linewidth]{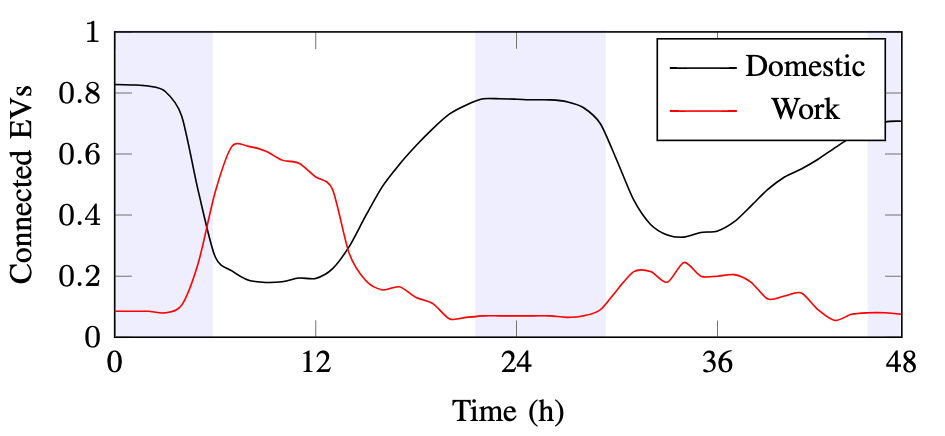}
\vspace*{-2mm}
\caption{Typical Fri-Sat normalised EV connection numbers for the `work' and `domestic' fleets. Nightime periods are shaded purple.}
\label{connection timeseries}
\vspace*{-1mm}
\end{figure}

A charging event records charge-point identity, EV connection and disconnection time. From these the continuous annual time-series of the number of connected EVs is derived. Fig.~\ref{connection timeseries} plots a typical Fri-Sat period for the two fleets. The domestic fleet is characterised by EVs disconnecting in the morning and then reconnecting in the afternoon. The work fleet exhibits the opposite trend during the week. Very few EVs connect to the work chargers on weekends.

The continuous time-series was discretized into average hourly connection and disconnection time series to provide $\Delta \hat{N}, \Delta N^{in}, \Delta N^{out}$, used in the constraints of Section \ref{SOC Section}. Average hourly values are appropriate for these constraints as they focus on net energy usage. 

\begin{figure}[!t]
\vspace*{-1mm}
\centering
\includegraphics[width=0.97\linewidth]{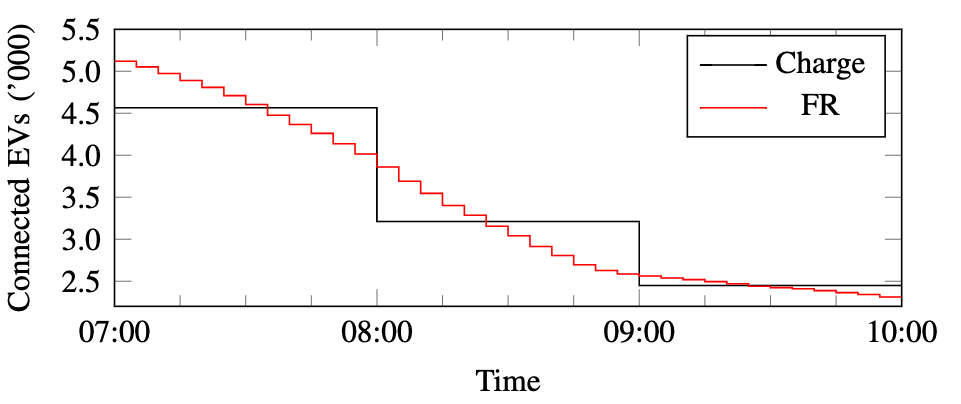}
\vspace*{-2mm}
\caption{Example of domestic EV connection values during a weekday morning with $N_0=5,119$, $t_d=$07:00 and $t_s$ = 07:00 $\rightarrow$ 08:00 or $t_s$ = 08:00 $\rightarrow$ 09:00 or $t_s$ = 09:00 $\rightarrow$ 10:00. FR must consider intra-hour connections (5-min) to define $\Delta N_i$, while charging decisions use average hourly values to define $\Delta \hat{N}$.}
\label{first hour}
\vspace*{-1mm}
\end{figure}

The need to consider intra-hour EV connections for FR scheduling is demonstrated in Fig.~\ref{first hour} which shows examples of the hourly and 5 minute discretized timeseries for the domestic fleet. EVs are disconnecting throughout the hours of the morning and the system operator wants to ensure the FR scheduled from EVs for each hour period is deliverable with 99\% security. Relying on the mean number of connected EVs to deliver FR would result in under delivery when there is an outage in the second half of the scheduling period. This risk to system dynamic security is unacceptable. Hence the 5-minute data is used when characterising $\Delta N$ for FR scheduling.

\begin{figure}[!t]
\centering
\includegraphics[width=0.97\linewidth]{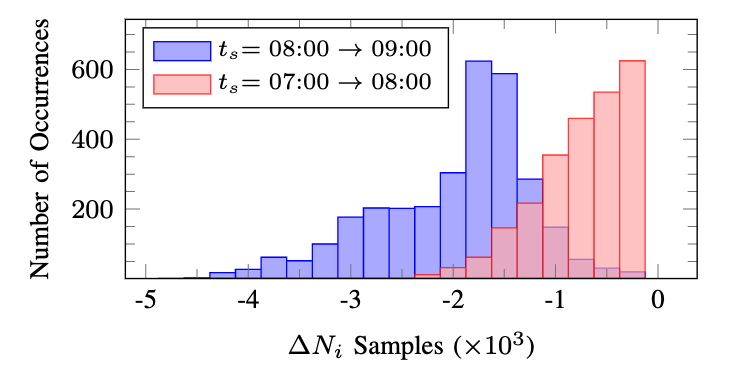}
\vspace*{-2mm}
\caption{Empirical distributions of $\Delta N_i$ for the domestic fleet on weekday, with $t_d=$07:00 and $t_s$ = 07:00 $\rightarrow$ 08:00 or $t_s$ = 08:00 $\rightarrow$ 09:00.}
\label{8am hist}
\end{figure}

\begin{figure}[!t]
\centering
\includegraphics[width=0.97\linewidth]{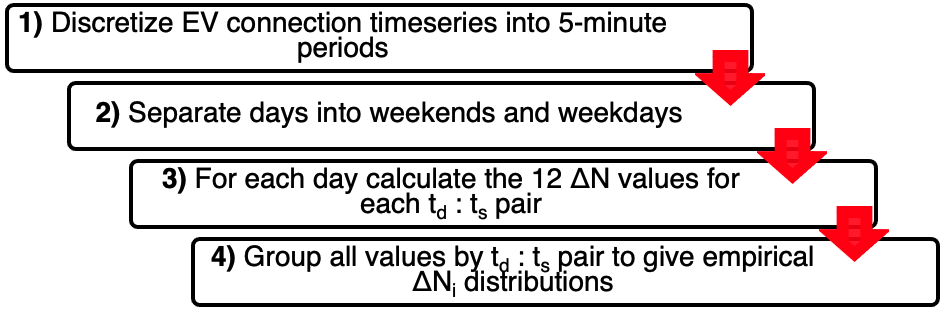}
\vspace*{-1mm}
\caption{Flow diagram to summarise forecasting approach. Empirical distributions for future intra-hour EV connectivity are assumed to be well characterised by driving behaviour from a similar time of the week the previous year. $t_d$ is the current time and $t_s$ is the future period in which FR from EVs is being scheduled.}
\label{Flow_Diagram}
\vspace*{-1mm}
\end{figure}

$\Delta N$ is the change in the number of connected EVs between the scheduling decision time ($t_d$) and during the scheduling period of interest ($t_s$). There are 260 weekdays and 105 weekend days in the 2017 data. This means that for a given $t_d$ on a weekday, there are $12 \cdot 260 = 3{,}140$ different samples of the possible change in the number of EVs between $t_d$ and during each $t_s$ period from the 5-minute discretized timeseries. Assuming all these are equiprobable, when collated together these 3,140 (or 1,260 for weekends) values produce an empirical distribution of $\Delta N$. Given that we are simplistically assuming here that the EV connection patterns do not significantly alter between years, then we can assume these historical empirical distributions are the true distributions of $\Delta N$ for use in scheduling FR in the future. Two example distributions for $t_d = 07:00$ and with $t_s = 07:00 \rightarrow 08:00, 08:00 \rightarrow 09:00$ are plotted in Fig. \ref{8am hist}. The 5-minute time-series from Fig.~\ref{first hour} contributes 12 data points to each histogram. The stages of this forecasting approach are summarised in Fig \ref{Flow_Diagram}.

\subsection{Tests for Ambiguity Set Selection}
The mean and standard deviation are necessary for all forms of (\ref{final form SOC}) described by Table \ref{f expressions}, but sufficient for only the most conservative DRO form. If more distributional information of $\boldsymbol{\delta}$ can be extracted from the distributions of its constituent $\Delta N_{i}$, then the ambiguity set can be tightened, and (\ref{final form SOC}) made less conservative. 

We can establish the likelihood that the $\Delta N_{i}$ distributions conform to unimodal or Gaussian distributions by applying the Shapiro-Wilk and Hartigans dip test respectively. Because $\boldsymbol{\delta}$ is the weighted sum of $\Delta N_{i}$ values, if they are established as Gaussian (and independent) with high likelyhood, then so too is $\boldsymbol{\delta}$. Indeed, if the $\Delta N_{i}$s can be shown to fit any distribution type with this summative property, then $f_P^{-1}$ can be accurately defined as the inverse cumulative distribution function. The sums of unimodal distributions are likely to be unimodal \cite{Roald2015}, but not necessarily unimodal. Despite the lack of this mathematical guarantee, we show empirically in Section \ref{Case Studies} that a unimodal assumption within (\ref{final form SOC}) can improve optimality whilst maintaining a significant conservative margin. Although to be guaranteed violation probabilities less than those specified, the DRO form must be implemented.

Table \ref{Stat tests Table} shows the results from applying the Shapiro-Wilk and Hartigans dip test to the 24 weekday and 24 weekend $\Delta N_{i}$ distribution where the scheduling period is the hour immediately following $t_d$ (i.e. $t_s = t_d \rightarrow t_d+1$). These are chosen because the operational scheduling model used here optimally schedules for the entire next 24hr period every hour. Thus, when it makes its scheduling decision, it is only the one for $t_s = t_d \rightarrow t_d+1$ that impacts reliability post fault, because the other decisions are subsequently revised to account for updated wind and EV connectivity realisations.

Typically the hypothesis (unimodality or normality) is rejected for p-values below 0.05 and accepted for values above 0.95. The hypothesis is neither rejected nor accepted for values between the two. Table \ref{Stat tests Table} shows that the Gaussian hypothesis is not accepted, obvious from Fig.~\ref{8am hist} due to the highly skewed distribution. On the other hand, 41/48 distributions are unimodal with high probability, with the other 7 being potentially unimodal.


\begin{table}[!t]
\renewcommand{\arraystretch}{1}
\caption{Tests on $\Delta N_i$ for Normality and Unimodality}
\label{Stat tests Table}
\centering
\begin{tabular}{|c|c|c|c|c|}
\hline P-Value &   $<$ 0.05 &  0.05 : 0.50 & 0.50 : 0.95 & $>$ 0.95\\
\hline
Normality & 19 & 12 & 12 & 5 \\
\hline
Unimodality & 0 & 0 & 7 & 41 \\
\hline
\end{tabular}
\vspace*{-3mm}
\end{table}


\section{Case Studies} \label{Case Studies}
 The uncertainty-aware model for V2G proposed in this paper was incorporated into the SUC model introduced in Section~\ref{SUC Section}. Case studies were run to identify how different EV fleet configurations and user risk-aversion impact the annual operational cost of the GB 2030 system. Unless otherwise stated, system thermal generation and storage mix was as listed in Table \ref{Genmix}.

An advanced statistical wind model \cite{Sturt2012,Sturt_conf} was used to characterise wind forecast uncertainty and produce a representative aggregate UK timeseries of wind realisations for the UK system in 2030. Full parametrization can be found in \cite{Sturt_conf}. The average load factor is 35\%. The time-series is normalised between 0 and 1 and then multiplied by the chosen GW of installed capacity, set at 40 GW unless otherwise stated. The aggregate UK solar generation time-series utilises the last year available for weather and system operator data \cite{PFENNINGER20161251}. Unless otherwise stated, 20~GW is installed with an average load factor of 11\%. The UK 2020 demand time-series was used to represent passive demand, whilst EV fleet demand is detailed in Section \ref{EV Connectivity Forecasting}. The annual passive demand ranges between 20:58 GW.

Generator and storage actions were optimally scheduled for 1 month of each season. A scenario tree that branches 7 times at the root node only was used to account for wind forecast uncertainty, which \cite{Sturt2012} showed to find the appropriate balance between tractability and optimality. Quantiles of 0.005, 0.1, 0.3, 0.5, 0.7, 0.9 and 0.995 were used.


Current GB frequency security standards were used: $f_0=50$~Hz, $|\Delta f_{max}|$ = 0.8~Hz and $RoCoF_{max}  = 1$~Hz/s.  The FR time constants are $T_1$ = 1s, $T_2$ = 10s, while $c_{LS}=$£30,000/MWh. Unless otherwise stated, two EV fleets were present on the system. With 85,000, 10~kW `Domestic' V2G chargers and 15,000, 20~kW `Work' V2G chargers. The parameters used were those derived in Section \ref{EV Connectivity Forecasting}, linearly scaled to match the total number. Nadir security was specified at 99\%.

An eight-core Intel Xeon 2.40GHz CPU with 64GB of RAM was used to run simulations. The optimisations were solved using XPRESS 8.12 linked to a C++ application via the BCL interface. The mixed-integer program gap was 0.1\%. 

\begin{table}[!t]
\caption{Generation and Storage Characteristics}
\label{Genmix}
\centering
\renewcommand{\arraystretch}{1}
\begin{tabular}{|l|l|l|l|}
\hline
\textbf{Generation} & Nuclear & CCGT & OCGT \\
\hline
Number of Units & 4 & 120 & 20 \\

Rated Power (GW) & 1.8 & 0.5 & 0.1 \\

Min Stable Generation (GW) & 1.60 & 0.25 & 0.05 \\

No-Load Cost (£'000/h) & 0.0 & 4.5 & 3.0\\

Marginal Cost (£/MWh) & 10 & 47 & 200 \\

Startup Cost (£'000) & NA & 10 & 0\\

Startup Time (h) & NA & 3 & 0 \\

Min up Time (h) & NA & 4 & 0 \\

Inertia Constant (s) & 5 & 4 & 4\\

Max Slow FR Capacity (GW) & 0.00 & 0.05 & 0.04\\
\hline
\hline
\textbf{Storage} & Pumped & Battery 1 & Battery 2 \\
\hline
Capacity (GWh) & 10 & 0.8 & 12 \\

Dis/Charge Rate (GW) & 2.6 & 0.4 & 3.0 \\

Max Fast FR Capacity (GW) & 0.0 & 0.8 & 0.0 \\

Max Slow FR Capacity (GW) & 0.5 & 0.0 & 0.0 \\

Dis/Charge Efficiency  & 0.75 & 0.95 & 0.95 \\
\hline
\end{tabular}
\vspace*{-3mm}
\end{table}

\subsection{Constraint Reliability}
\begin{figure}[!t]
\centering
\includegraphics[width=0.97\linewidth]{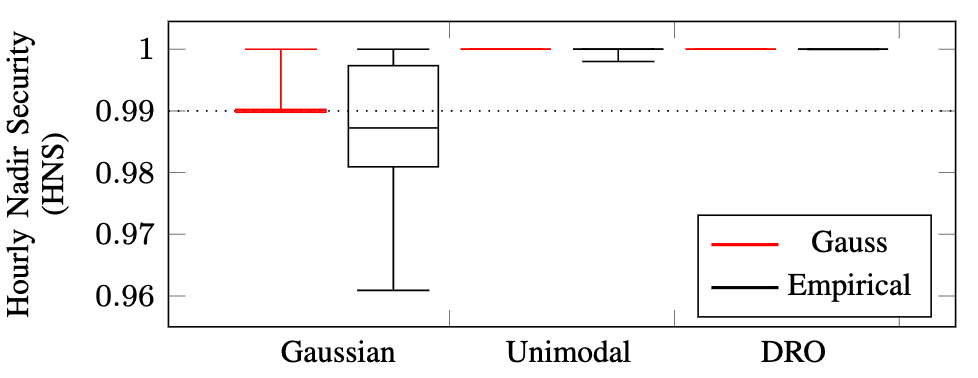}
\vspace*{-2mm}
\caption{ The inter-quartile range, median and max/min of hourly nadir security levels under different ambiguity set assumptions (Gaussian, Unimodal, DRO) on $\boldsymbol{\delta}$ when scheduling FR via (\ref{final form SOC}). The scheduled amount was compared to true EV connectivity, found by sampling Gaussian or Empirical distributions of $\Delta N_i$.}
\label{deliverability prob 1}
\end{figure}

Fig.~\ref{deliverability prob 1} demonstrates how the specified risk of $\boldsymbol{\bar{R}^{EV}}$ under delivery ($\epsilon$) compares to actual deliverability when using (\ref{final form SOC}) to schedule FR from EVs during annual SUC simulations. This varies depending on the ambiguity set assumptions (Gaussian, unimodal or DRO) and the assumed true distribution of $\Delta N_i$ (Gaussian or empirical). Hourly nadir security (HNS) is the metric used to evaluate $\boldsymbol{\bar{R}^{EV}}$ deliverability. HNS for a specific hour is found by sampling the $\Delta N_i$ distribution for each fleet. When added to the number of currently connected EVs, the actual FR deliverable ($R^{EV}_j$) if an outage occurred at a random time over the scheduling period can be calculated. This process is repeated 100,000 times within each hour. The HNS is the cumulative ratio of $\boldsymbol{\bar{R}^{EV}} > R^{EV}_j$. Fig.~\ref{deliverability prob 1} plots the range of HNS for different constraints over the simulation period. It only shows the $t_s$ in the hour immediately following $t_d$. Due to the rolling planning approach of the SUC, the $\boldsymbol{\bar{R}^{EV}}$ values for other $t_s$ are revised before the system would experience an outage. 

When $\boldsymbol{\delta_n}$ is assumed Gaussian and the $R^{EV}_j$ is found from sampling Gaussian distributed $\Delta N_i$, the HNS exactly equals the specified security level of 99\% when constraint (\ref{final form SOC}) is tight. The constraint is occasionally not tight during periods of high net demand when inertia and slow FR from thermal plants are sufficient to meet frequency security needs. During these times the HNS takes values above 99\%. However, when the true empirical $\Delta N_i$ distributions are sampled instead, 50\% of hours have a HNS less than the specified 99\%. In the worst period, using the Gaussian (\ref{final form SOC}) constraint would result in only 96\% of outages being contained securely. This unreliability is in line with the analysis of Table \ref{Stat tests Table}, which showed none of the relevant empirical $\Delta N_i$ distributions are Gaussian. Thus using the Gaussian form of (\ref{final form SOC}) is over optimistic and risks system security, so is not considered further.

On the other hand, Fig. \ref{deliverability prob 1} shows that applying  (\ref{final form SOC}) assuming a unimodal or DRO $\boldsymbol{\delta_n}$ distribution results in conservative HNS values for both the Gaussian and empirical distribution sampling. For Unimodal and DRO the worst periods have a HNS of 99.7\% 99.9\% respectively, with the majority of hours delivering more response than is scheduled with close to 100\% probability. Again, this is in line with the analysis of Table \ref{Stat tests Table}, which showed a high likelihood that $\Delta N$s are unimodal. 



\subsection{Value of Response from V2G and Comparison to State of the Art}

A scheduling method's success criteria is to schedule as much FR from EVs as possible to improve operational efficiency whilst ensuring that it is deliverable at the specified certainty level or higher. Table \ref{results table} compares the value of FR from EVs when using our proposed `joint' method (\ref{final form SOC}) to the value when using three simple methods that are the current state of the art. It shows it to be more secure than the deterministic approach whilst less conservative than the `No V2G' and `individual' approach.

The `No V2G FR' row in Table \ref{results table} allows no response from distributed resources. Zero FR is deliverable 100\% of the time but offers no operational cost savings. The value of FR from EVs using the other methods is defined as the reduction in system annual operational costs compared to `No V2G FR'. The value creation mechanisms are explored in Section \ref{time-varying contribution}.

A system operator could schedule FR using a simple deterministic approach, where it is assumed that the mean forecast number of EV connections will occur. This can be done by making `$\sigma_i = 0 \ \forall \ i$' in (\ref{final form SOC}), resulting in a linear constraint. By discounting uncertainty in this way, larger amounts of $\boldsymbol{\bar{R}^{EV}}$ are scheduled thus annual system costs are significantly reduced by £8,040/charger. However, because the mean is used, half of the time fewer EVs are available to provide response than is scheduled for. Thus this cost reduction comes at an unacceptable loss of nadir security, with average and worst case HNS being only 50\%. 

\begin{table}[!t]
\renewcommand{\arraystretch}{1}
\caption{Constraint Value Comparison for Specified 99\% Frequency Security and 100,000 Chargers}
\centering
\label{results table}
\begin{tabular}{|c|c|c|c|}
\hline Constraint &   \begin{tabular}{@{}c@{}} Worst Case \\ Nadir Security \end{tabular} &  \begin{tabular}{@{}c@{}} Charger \\ Value (£/yr) \end{tabular} & \begin{tabular}{@{}c@{}} Charger CO$_2$ \\ Savings (tons/yr) \end{tabular}\\
\hline
\hline
No V2G FR & 100.0 & 0  & 0 \\
\hline
Deterministic & 50.0 & 8,040  & 44.5 \\
\hline
Unimodal (\ref{Indv 3}) & 100.0 & 5,930  & 34.1 \\
\hline
Unimodal (\ref{final form SOC}) & 99.7 & 6,330  & 37.4 \\
\hline
DRO (\ref{Indv 3}) & 100.0 & 5,200  & 31.7 \\
\hline
DRO (\ref{final form SOC}) & 99.9 & 5,760  & 33.6 \\
\hline
\end{tabular}
\vspace*{-3mm}
\end{table}

EV connection uncertainty can be considered simplistically via the `individual' method (\ref{Indv 3}). As demonstrated in (\ref{Indv 4}), our more advanced `joint' approach (\ref{final form SOC}) will always schedule more response than (\ref{Indv 3}), thus system operation is improved and a charger's value is increased by 6.9\% and 10.9\% for unimodal and DRO ambiguity set assumptions respectively. Crucially this comes with worst case HNS being maintained well above the specified 99\% security requirement.

\begin{figure}[!t]
\centering
\includegraphics[width=0.97\linewidth]{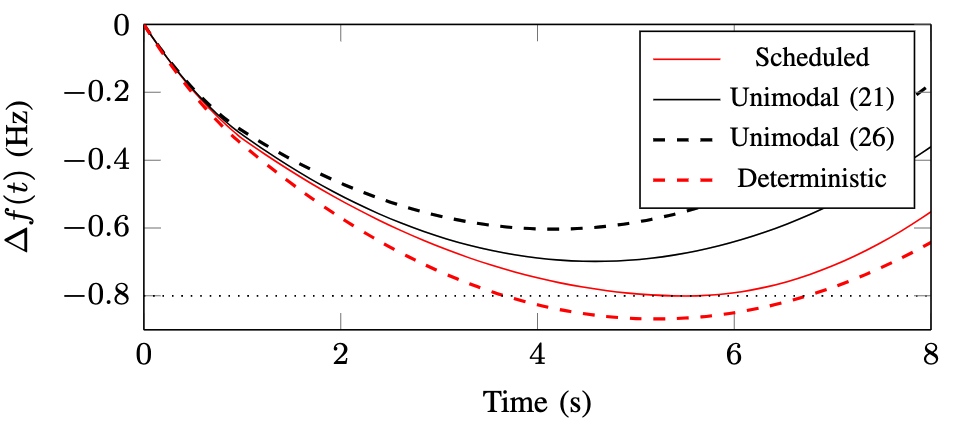}
\caption{Comparison of the frequency evolution post $\boldsymbol{PL_{max}}$ loss given the delivery of the scheduled amount of FR  $\boldsymbol{\bar{R}^{EV}}$ or of the 1st percentile of the true FR distribution $\boldsymbol{R^{EV}}$ for the system conditions shown in Table \ref{Dynamic Sim table}.  The `scheduled' plots are very similar so for clarity only one is plotted.}
\label{Dynamic sims}
\end{figure}

For 99\% of outages the EVs deliver more FR than the first percentile of $\boldsymbol{R^{EV}}$ ($R^{EV}_{1st}$). Specifiying $\epsilon = 0.01$ is equivalent to requiring that the frequency nadir remains higher than 49.2Hz when $R^{EV}_{1st}$ GW is delivered. Table \ref{Dynamic Sim table} lists the frequency services scheduled for a typical weekday 09:00 period under four scheduling methods. In whilst Fig.~\ref{Dynamic sims} a time-domain simulation in MATLAB was used to plot their dynamic frequency evolution when $R^{EV}_{1st}$ is delivered instead of the scheduled $\boldsymbol{R^{EV}}$. 

When the scheduled amount of $\boldsymbol{\bar{R}^{EV}}$ is delivered, the constraint (\ref{Nadir Constraint}) is tight so the nadir is exactly 49.2~Hz. A scheduling method with perfect EV connection forecasts would emulate this when $R^{EV}_{1st}$~GW of FR is delivered. Fig~\ref{Dynamic sims} demonstrates that our `joint' method is closer to this ideal than the `individual' approach, that significantly under schedules FR from EVs. This is undesirable as being overly conservative inflates operational costs. The conservativness of the `joint' approach is attributed to its robustness in covering the entire unimodal ambiguity set. On the other hand, the deterministic method is overly optimistic and over schedules FR from EVs. Thus 50\% of the time the frequency breaches the nadir limit, jeopardising system security and demonstrating the importance of applying a risk-aware scheduling method like the ones derived in this paper. 

\begin{table}[!t]
\renewcommand{\arraystretch}{1.2}
\caption{Dynamic Simulation Parameters}
\label{Dynamic Sim table}
\centering
\begin{tabular}{|c||c|c|c|c|c|c|}
 \cline{2-7}
\multicolumn{1}{c||}{} & 
\hspace{-0.2cm} \begin{tabular}{@{}c@{}} $ \boldsymbol{\bar{R}^{EV}}$ \\ (GW) \end{tabular}&
\hspace{-0.2cm} \begin{tabular}{@{}c@{}} $ R^{EV}_{1st}$ \\ (GW) \end{tabular}&
\hspace{-0.2cm} \begin{tabular}{@{}c@{}} $ \boldsymbol{R^{ND}}$ \\ \footnotesize{(GW)} \end{tabular}&
\hspace{-0.2cm} \begin{tabular}{@{}c@{}} $ \boldsymbol{R^G}$ \\ \footnotesize{(GW)} \end{tabular}&
\hspace{-0.2cm} \begin{tabular}{@{}c@{}} $ \boldsymbol{H}$ \\ \footnotesize{(GWs)} \end{tabular}  &
\hspace{-0.2cm} \begin{tabular}{@{}c@{}} $ \boldsymbol{PL_{\small{max}}}$ \\ \footnotesize{(GW)} \end{tabular} \hspace{-0.2cm} \\
\hline
\hline
\begin{tabular}{@{}c@{}}\footnotesize{Unimodal} \\ \footnotesize{(\ref{final form SOC})} \end{tabular} & 0.23 & 0.33 & 0.36 & 2.27 & 96.25 & 1.73\\
\hline
\begin{tabular}{@{}c@{}}\footnotesize{Unimodal} \\ \footnotesize{(\ref{Indv 3})} \end{tabular} & 0.22 & 0.42 & 0.40 & 2.36 & 101.49 & 1.80\\
\hline
\begin{tabular}{@{}c@{}}\footnotesize{Determ} \vspace{0.25cm} \\  \end{tabular} & 0.28 & 0.22 & 0.40 & 2.23 & 96.24 & 1.78\\
\hline
\end{tabular}
\end{table}
\renewcommand{\arraystretch}{1}

\subsection{Constraint Performance with Varied Fleet Types}

A strength of our proposed formulation is its wide applicability to systems with any number and type of fleet so long as their $\sigma_i$ and $\mu_i$ are known. Accordingly, Table \ref{Joint vs Indv} shows the improvement, in value and amount of scheduled FR, of the `joint' approach over the `individual' approach for a range of fleet setups.

According to (\ref{Indv 4}), the improvement of our `joint' method over the state of the art increases with the ratio of $\sum_{i \in I} \boldsymbol{g_i} \delta_i : \sqrt{\sum_{i \in I} (\boldsymbol{g_i} \delta_i)^2}$. This ratio will increase when $ \boldsymbol{g_i} \delta_i$ are of a similar size, explaining the increased improvement of our method for the system with two work fleets of 15,000 EV,  as opposed to the case with one `Domestic' fleet with 85,000 EVs and one `Work' fleet. Our method will also increasingly outperform the current state of the art as the number of fleets increases, demonstrated by comparing the cases with two and three `Work' fleets.

\begin{table}[!t]
\renewcommand{\arraystretch}{1}
\caption{Increase of FR from EVs Using (\ref{final form SOC}) over (\ref{Indv 3})}
\centering
\label{Joint vs Indv}
\begin{tabular}{|c|c||c|c||c|c|}
\hline \multicolumn{1}{|c}{Fleets} &  &   \multicolumn{1}{|c}{Unimod}  &   & \multicolumn{1}{|c}{DRO} & \\
\hline
Domestic & Work & Value (£) & FR (GW) & Value & FR \\
\hline
\hline
1 & 0 & 0.0\% & 0.0\% & 0.0\% & 0.0\% \\
\hline
1 & 1 & 6.9\% & 5.4\% & 10.9\% & 8.5\% \\
\hline
2 & 0 & 28.0\% & 22.6\% & 49.2\% & 40.3\% \\
\hline
3 & 0 & 38.6\% & 32.6\% & 70.1\% & 61.0\% \\
\hline
\end{tabular}
\end{table}

\begin{figure}[!t]
\centering
\includegraphics[width=0.97\linewidth]{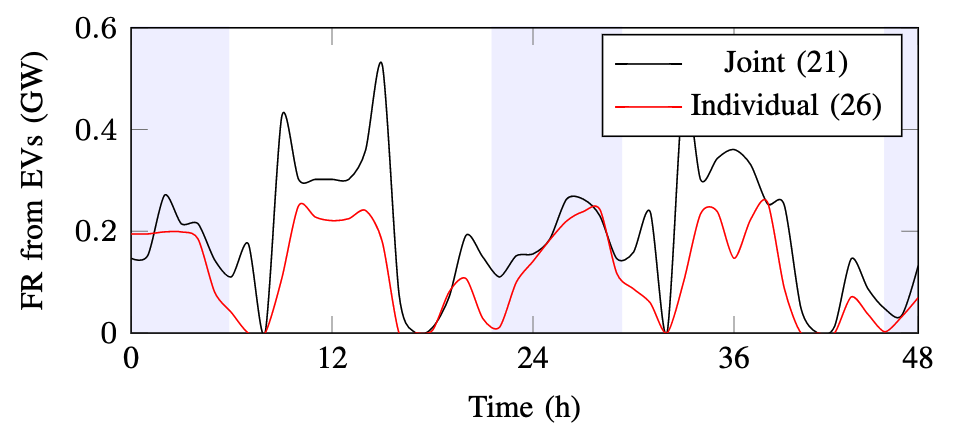}
\vspace*{-1mm}
\caption{Total scheduled FR from three 15,000 EV `work' fleets over the same two-day period, using the `joint' (\ref{final form SOC}) and `individual' (\ref{Indv 3}) approaches. Nighttime is shaded purple.}
\label{FR Timeseries}
\end{figure}

Fig.~\ref{FR Timeseries} demonstrates the increased conservativness of the current state of the art over our proposed method by plotting the total $\boldsymbol{\bar{R}^{EV}}$ over the same two-day period using both methods for a case with three `Work' fleets with 15,000 EV each, under DRO ambiguity assumptions. During the highly uncertain period in the morning and evening when EVs are arriving and departing at work, neither method can reliably schedule any FR from the fleets. However, at other times the `joint' method is able to schedule more FR than the individual, with up to 0.38~GW more in the period shown. More FR from EVs translates into reduced operational costs, thus increased charger value.


\begin{figure}[!t]
\centering
\includegraphics[width=0.97\linewidth]{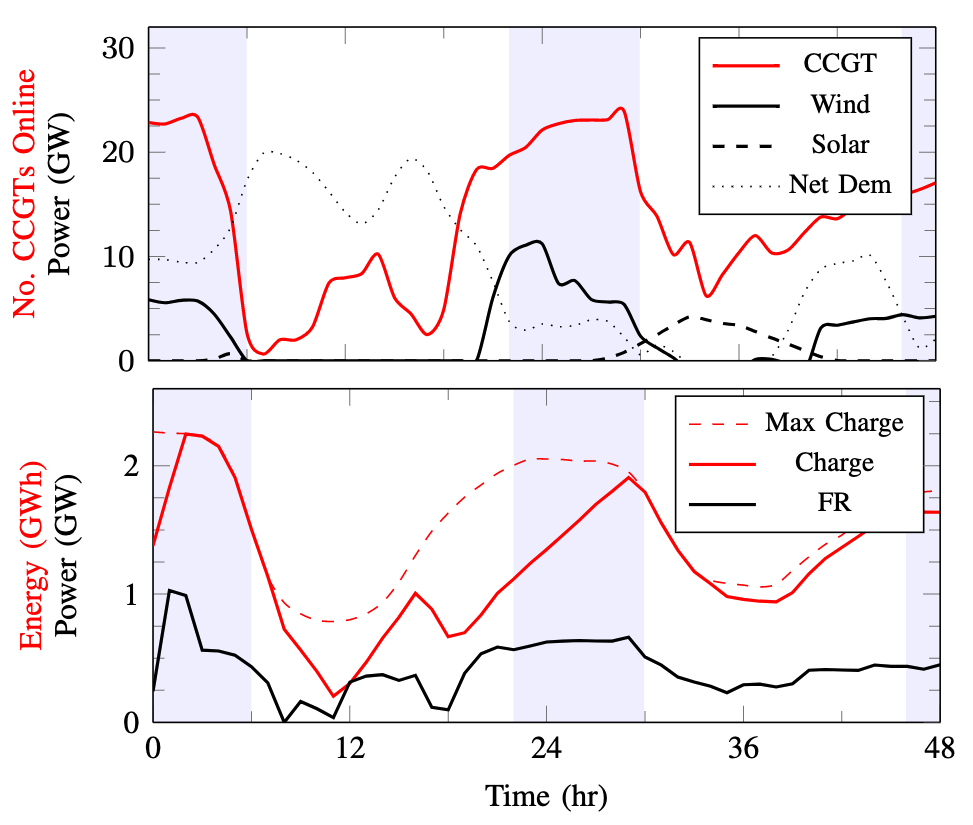}
\vspace*{-2mm}
\caption{Two day example comparing system operation with and without FR from V2G. Nighttime is shaded purple. The top figure plots the difference in: CCGT commitment number; wind and solar curtailment between the two systems. The bottom figure plots the operation of the Domestic fleet virtual battery for the system with V2G FR enabled.}
\label{WindShed Timeseries}
\vspace*{-2mm}
\end{figure}

\subsection{Investigating time-varying FR contribution from EVs} \label{time-varying contribution}
Fig.~\ref{WindShed Timeseries} shows how FR provision from aggregated EV fleets creates system value by significantly reducing wind and solar curtailment during low net-demand periods. It compares the operating conditions of the system with and without FR from EVs enabled over a typical two-day period. The unimodal constraint is used. The net difference in committed CCGTs, wind and solar curtailment are plotted. Net demand is also plotted, this is equal to the total wind and solar energy available subtracted from demand, so is the same for both systems. The crucial difference is in how much more of this available zero marginal cost and emissions-free renewable power the system with FR from V2G is able to integrate.

During the first day, net demand is above 10 GW. The inertia from CCGTs committed to serve this load in combination with system FR is enough to fully secure frequency so no wind shed occurs in either system. However, during the second night demand drops and wind energy increases, resulting in low (and even negative) net-demand for the second day. In the system without FR from V2G, the inertia from thermal plants needed for energy provision alone is insufficient to secure the nadir~(\ref{Nadir Constraint}). Consequently CCGTs must be committed for their inertia and FR alone. When the sum of the minimum stable generation of online thermal plants is larger than net demand, renewable power must be curtailed to respect the power balance constraint (\ref{power balance}). The clear correlation between over commitment of CCGTs and wind curtailment is shown in Fig.~\ref{WindShed Timeseries},  when during the second night around 24 more CCGTs are online for the system without V2G FR. Thus the sum of their 250MW individual minimum stable generation results in roughly 6GW wind curtailment. 

Fig.~\ref{WindShed Timeseries} also plots the operation and FR provision of the domestic fleet during the same two-day period. During the second night the EVs are charged at a constant rate. The need for full charge by morning synergises with the typically increased FR value overnight caused by lower net-demand. FR is a net power injection thus a charging EV can provide more response via demand alleviation. The approximately 0.6~GW of FR from the fleet during the second night replaces the inertia from approximately 20 additional CCGTs, facilitating up to 11GW of wind integration. Cumulatively over the year the net difference in wind and solar curtailment amounts to 7.55~TWh and 0.40~TWh respectively. In other words, the highly effective FR from V2G connected EVs enables frequency secure operation at very low inertia levels, resulting in 8~TWh less power generated by burning fossil fuels (80~MWh per charger). This accounts for the majority of cost and emission savings in Table~\ref{results table}.

\subsection{Value's Sensitivity to User Risk aversion and Forecast Uncertainty}

\begin{figure}[!t]
\centering
\includegraphics[width=0.97\linewidth]{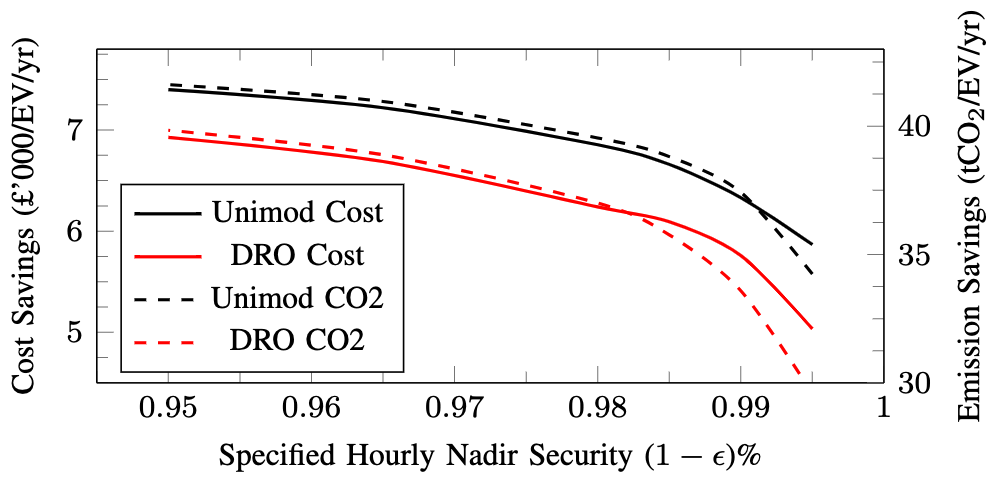}
\vspace*{-2mm}
\caption{Sensitivity of charger value to FR under-delivery risk aversion.}
\label{Value per EV}
\vspace*{-1mm}
\end{figure}

A useful feature of (\ref{final form SOC}) is that it directly translates a user's risk aversion level ($\epsilon$) into a scheduled amount of FR. Fig.~\ref{Value per EV} illustrates that higher risk aversion leads to less response allowed from EVs and thus lowers their value. The tightened ambiguity set from assuming $\boldsymbol{\delta_n}$ is unimodal results in a larger $\frac{1}{ f^{-1}_\mathcal{P}(1-\epsilon)}$ constant in (\ref{final form SOC}), and thus more $\boldsymbol{\bar{R}^{EV}}$ can be scheduled than the DRO case for the same $\epsilon$. This explains the increased operation cost savings shown in both Table~\ref{results table} and Fig.~\ref{Value per EV}.

\begin{figure}[!t]
\centering
\includegraphics[width=0.97\linewidth]{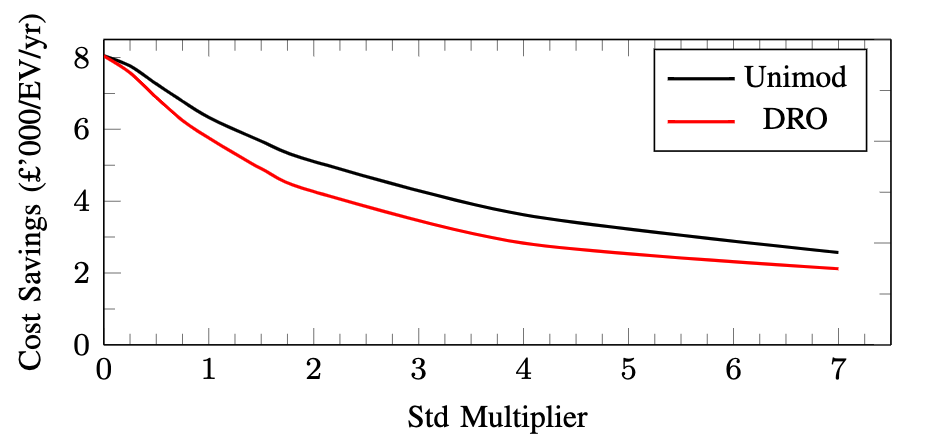}
\vspace*{-2mm}
\caption{Sensitivity of charger value to increased standard deviation (std) of $\Delta N_i$.}
\label{Std Sensitivity}
\vspace*{-3mm}
\end{figure}

A large $\sigma_i$ of $\Delta N_i$ implies that the number of connected EVs in the scheduling period of interest is highly variable. This decreases the amount of schedulable FR from the chargers because a high deliverability probability necessitates covering the edge cases where far fewer EVs than the mean are connected at the time of an outage. A reduced std has the opposite effect, increasing the amount of schedulable FR, this is reflected in Fig. \ref{Std Sensitivity} which shows how the V2G's value varies in relation to a multiplier applied to the stds in equation (\ref{final form SOC}). As the stds tend to zero, the cost saving tends towards the deterministic case level of £8,000/yr, with the difference between DRO and unimodal ambiguity set assumptions diminishing. It is interesting to observe that even with very large stds the EVs still provide substantial value of approximately £2,000/yr, because their connectivity variability overnight is very low, so FR provision during this valuable period is mostly uninterrupted.

The relationship between a small std and increased V2G value has two main implications for real life application of (\ref{final form SOC}). Firstly, more accurate EV connectivity forecasting methods are directly incentivised. Secondly, closer to real time scheduling of FR is desirable, as it reduces uncertainty in forecasting.

\subsection{Value's Sensitivity to System Characteristics}
\begin{figure}[!t]
\centering
\includegraphics[width=0.97\linewidth]{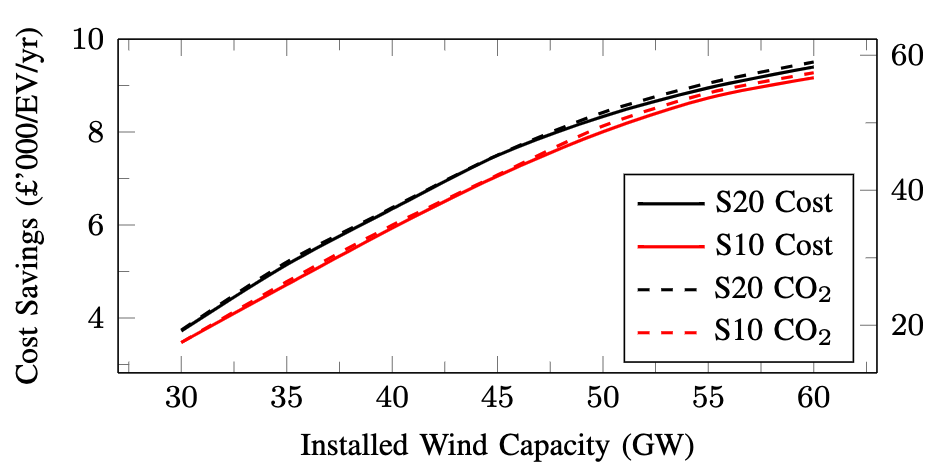}
\vspace*{-2mm}
\caption{The main value creation mechanism of FR from V2G chargers is the ability to facilitate renewable generation integration by displacing inertia and FR from thermal plants. The value sensitivity to installed system wind capacity and 10GW (`S10') and 20GW (`S20') of solar capacity is plotted here.}
\label{Renewable Penetration}
\vspace*{-3mm}
\end{figure}

The value of response from V2G is highly dependant on its ability to facilitate higher renewable integration by displacing the inertia and FR from thermal plants. Consequently, Fig.~\ref{Renewable Penetration} shows that increased renewable generation increases V2G value. With 60~GW of wind and 20~GW solar, one V2G charger has an annual system value of £9,400/yr. This comes from the increased frequency and magnitude of low-inertia periods, where the FR from the 100,000 chargers facilitate a cumulative 14.3 TWh of renewable power integration.

\begin{figure}[!t]
\centering
\includegraphics[width=0.97\linewidth]{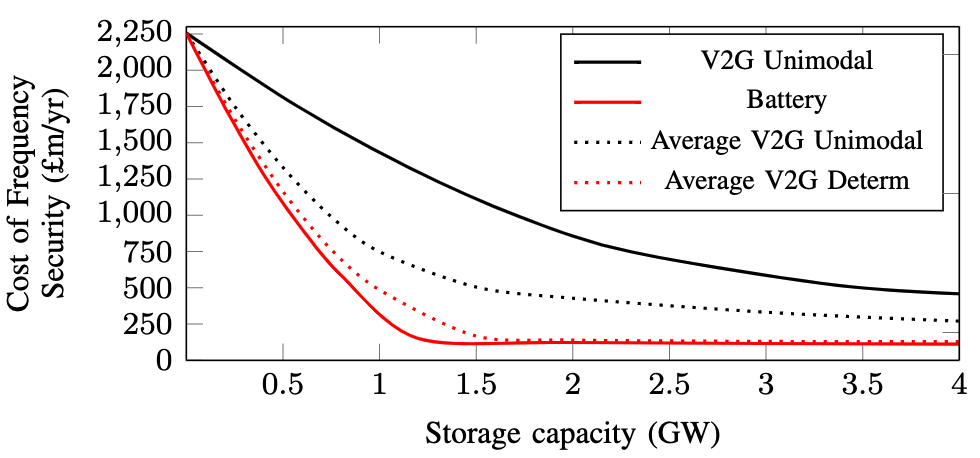}
\vspace*{-1mm}
\caption{Operational cost increase from applying the frequency security constraints to systems with different storage penetrations. The difference between battery and V2G is primarily because the chargers on average only have an EV connected (thus FR capability) $\approx$40\% of the time. Adjusting for this gives the `Average V2G' capacity plots.}
\label{Bat_pen}
\vspace*{-1mm}
\end{figure}

The cost of frequency security (CFS) is the operational cost increase incurred by applying the nadir and RoCoF constraint. We now analyse the CFS's dependence on the amount of storage on the system. As discussed in Section \ref{time-varying contribution} this cost increase comes from the need to run CCGTs to provide inertia and FR to satisfy the nadir and RoCoF constraints. Running CCGTs burns fuel at a cost, and often their cumulative minimum stable generation forces the curtailement of zero marginal cost and emission-free renewable energy. In systems with high renewable penetrations this cost increase is very significant. Fig. \ref{Bat_pen} shows that it reaches £2.26bn/yr (22\% of total operational cost) for our standard test system of 20GW solar and 40GW wind, when neither V2G or batteries can provide FR.

FR from inverter-based resources reduces the required inertia and FR from thermal plants and thus increases renewable integration and reduces the CFS. A core strength of our proposed formulation is that it allows the abundant distributed resources that will be present in future systems (like V2G-connected EVs) to compete directly with grid batteries to provide this FR, whilst maintaining a user specified guarantee on system dynamic security. Operational costs for batteries and V2G are assumed zero so do not contribute to CFS.

Fig.~\ref{Bat_pen} shows that the first 2.25~GW of V2G capacity is a third as effective at reducing the CFS as battery capacity. For example, to reduce FSC to £1.5bn/yr requires 0.35~GW of batteries or 0.95~GW of V2G (66,500 `Domestic' and 14,250 `Work' chargers). Reduction to £0.75bn/yr requires 0.75~GW of batteries or 2.30~GW of V2G (161,000 `Domestic' and 34,500 `Work' chargers).  Above 2.50~GW of V2G capacity, FR is abundant enough that renewable shedding only occurs during the highly uncertain morning period. During these periods the $\sigma_i$ values are large enough that $\bar{R}^{EV}$ in (\ref{final form SOC}) is forced to be very small or zero, as occurs during the two morning periods in Fig. \ref{FR Timeseries}. Increased V2G capacity alleviates this slowly, explaining the plateau in CFS reduction from V2G capacity. Whereas battery storage has no uncertainty so the same effect is not observed. However, the marginal value of storage does saturate above 1.2 GW, at which point the minimum inertia for the RoCoF constraint (to which FR does not contribute) dominates the CFS.

The lower value of FR from V2Gs compared to the same capacity of batteries is primarily attributable to a V2G charger only offering FR when an EV is connected. From the fleet parameters derived in Section \ref{EV Connectivity Forecasting}, the average charger has an EV plugged in 42\% and 26\% of the time for `Domestic' and `Work' respectively. This is adjusted for within Fig. \ref{Bat_pen} with the average annual V2G capacity available plotted against value created for both the deterministic and unimodal constraints. The small difference between deterministic and the battery capacity is attributable to the EVs time of connection and charge requirements to meet energy needs. The difference between the unimodal and deterministic plots is due to uncertainty, revealing this to be the second most impactful derating factor. Uncertainty has low impact below 1~GW of average capacity, but above this it prohibits frequency security cost reduction again due to the highly uncertain morning periods. This suggests that the addition of a fleet with low morning plugin uncertainty, or a small amount of grid batteries would be valuable at high EV penetrations.

\begin{table}[!t]
\renewcommand{\arraystretch}{1}
\caption{Sensitivity of Normalised V2G Value to FR Provision Delay}
\label{Delays}
\centering
\begin{tabular}{|c|c|c|c|c|c|c|}
\hline Delay (s) & 0 & 0.2 &  0.4 & 0.6 & 0.8 & 1.0 \\
\hline
Normalised Value & 1.00 & 0.97 & 0.93  & 0.89 & 0.85 & 0.80 \\
\hline

\end{tabular}
\vspace*{-3mm}
\end{table}

\subsection{Impact of Delays on Value of EV's Response}
All other sections of this paper assume FR from EVs begin ramping up the instant of $\boldsymbol{PL_{max}}$ disconnection. In reality, the triggering mechanism may involve delays caused by communication or frequency measurement requirements. The additional term in (\ref{Delay definition}) allows the impact of delays on the value of FR from V2G connected EVs to be analysed within the SUC. The results are shown in Table \ref{Delays}. The normalised value decrease is shown, this is the same for both the unimodal and DRO cases. Delays decrease the efficacy of response in containing frequency drop and thus reduce its value. Although, even when the final FR delivery time is doubled with a 1s delay, the value of FR remains substantial at £5,060/EV/yr for the unimodal case, or a decrease of 20\% with respect to the case with no delay.


\section{Conclusion and Future Work} \label{conclusions}

This paper proposes a mathematical framework to schedule frequency response from aggregated V2G chargers under uncertainty in EV plug-in times.
A distributionally-robust chance constrained formulation for the frequency-security limits is introduced, compatible with any probabilistic forecast for EV connections and providing a convex feasible region. Case studies demonstrate that the proposed scheduling methodology facilitates large cost savings ($\approx$ £6,000/charger) in the future Great Britain low inertia system due to displaced inertia and FR requirements from thermal plants. Crucially, this value is obtained with mathematical guarantees on system frequency security.

In future, a model that accounts for charger location should be developed. Given the significant economic benefits that V2G could provide, it will be key to coordinate V2G power injections to ensure that distribution-side network constraints are respected. Secondly, the design of a market for FR that allows aggregator participation should be investigated. The tool developed here allows market clearing under security guarantees, but the mechanism of specifying, communicating and monitoring aggregator uncertainty characteristics needs development.

\section*{Acknowledgment}
This research has been supported by the UK EPSRC project `Integrated Development of Low-Carbon Energy Systems' (IDLES, Grant EP/R045518/1), and by the Innovate UK project `e4Future' (104227).

\ifCLASSOPTIONcaptionsoff
  \newpage
\fi

\vspace*{-2mm}
\IEEEtriggeratref{3}

\bibliographystyle{IEEEtran}
\bibliography{Bibliography.bib}

\begin{thebibliography}{10}
\providecommand{\url}[1]{#1}
\csname url@samestyle\endcsname
\providecommand{\newblock}{\relax}
\providecommand{\bibinfo}[2]{#2}
\providecommand{\BIBentrySTDinterwordspacing}{\spaceskip=0pt\relax}
\providecommand{\BIBentryALTinterwordstretchfactor}{4}
\providecommand{\BIBentryALTinterwordspacing}{\spaceskip=\fontdimen2\font plus
\BIBentryALTinterwordstretchfactor\fontdimen3\font minus
  \fontdimen4\font\relax}
\providecommand{\BIBforeignlanguage}[2]{{%
\expandafter\ifx\csname l@#1\endcsname\relax
\typeout{** WARNING: IEEEtran.bst: No hyphenation pattern has been}%
\typeout{** loaded for the language `#1'. Using the pattern for}%
\typeout{** the default language instead.}%
\else
\language=\csname l@#1\endcsname
\fi
#2}}
\providecommand{\BIBdecl}{\relax}
\BIBdecl

\bibitem{Teng2016}
F.~Teng \emph{et~al.}, ``{Stochastic Scheduling with Inertia-Dependent Fast
  Frequency Response Requirements},'' \emph{IEEE Transactions on Power
  Systems}, vol.~31, no.~2, pp. 1557--1566, mar 2016.

\bibitem{Badesa2019}
L.~Badesa \emph{et~al.}, ``{Simultaneous Scheduling of Multiple Frequency
  Services in Stochastic Unit Commitment},'' \emph{IEEE Transactions on Power
  Systems}, vol.~34, no.~5, pp. 3858--3868, 2019.

\bibitem{ClimateChangeCommittee2020}
{Climate Change Committee}, ``{The UK's transition to electric vehicles},''
  2020.

\bibitem{OMalley2020}
C.~O’Malley \emph{et~al.}, ``Value of fleet vehicle to grid in providing
  transmission system operator services,'' in \emph{2020 Fifteenth
  International Conference on Ecological Vehicles and Renewable Energies},
  2020.

\bibitem{Thingvad2019}
A.~Thingvad \emph{et~al.}, ``Value of {V2G} frequency regulation in {Great}
  {Britain} considering real driving data,'' in \emph{2019 IEEE PES Innovative
  Smart Grid Technologies Europe (ISGT-Europe 2019)}, 2019.

\bibitem{Gao2021}
S.~Gao \emph{et~al.}, ``{Optimal Charging of Electric Vehicle Aggregations
  Participating in Energy and Ancillary Service Markets},'' \emph{IEEE Journal
  of Emerging and Selected Topics in Industrial Electronics}, vol.~3, no.~2,
  pp. 270--278, aug 2021.

\bibitem{BLATIAK2022100738}
A.~Blatiak \emph{et~al.}, ``{Value of optimal trip and charging scheduling of
  commercial electric vehicle fleets with Vehicle-to-Grid in future low inertia
  systems},'' \emph{Sustainable Energy, Grids and Networks}, vol.~31, p.
  100738, 2022.

\bibitem{Hajebrahimi2020}
A.~Hajebrahimi \emph{et~al.}, ``Scenario-wise distributionally robust
  optimization for collaborative intermittent resources and electric vehicle
  aggregator bidding strategy,'' \emph{IEEE Transactions on Power Systems},
  vol.~35, pp. 3706--3718, 2020.

\bibitem{Lu2020}
X.~Lu \emph{et~al.}, ``A model to mitigate forecast uncertainties in
  distribution systems using the temporal flexibility of {EVAs},'' \emph{IEEE
  Transactions on Power Systems}, vol.~35, pp. 2212--2221, 5 2020.

\bibitem{Amini2020}
M.~Amini and M.~Almassalkhi, ``{Optimal Corrective Dispatch of Uncertain
  Virtual Energy Storage Systems},'' \emph{IEEE Transactions on Smart Grid},
  vol.~11, no.~5, pp. 4155--4166, 2020.

\bibitem{NationalGridESO2022}
{National Grid ESO}, ``{Reintroduction of aggregation at GSP Group for DC},''
  {January} 2022.

\bibitem{Zhang2017}
Y.~Zhang \emph{et~al.}, ``Distributionally robust chance-constrained optimal
  power flow with uncertain renewables and uncertain reserves provided by
  loads,'' \emph{IEEE Transactions on Power Systems}, vol.~32, pp. 1378--1388,
  2017.

\bibitem{Bagchi2021}
A.~Bagchi \emph{et~al.}, ``Investigating impacts of storage devices on
  distribution network aggregator's day-ahead bidding strategy considering
  uncertainties,'' \emph{IEEE Access}, vol.~9, pp. 120\,940--120\,954, 2021.

\bibitem{Roald2015}
\BIBentryALTinterwordspacing
L.~Roald \emph{et~al.}, ``{Security Constrained Optimal Power Flow with
  Distributionally Robust Chance Constraints},'' pp. 1--8, 2015. [Online].
  Available: \url{http://arxiv.org/abs/1508.06061}
\BIBentrySTDinterwordspacing

\bibitem{Xie2018}
W.~Xie and S.~Ahmed, ``Distributionally robust chance constrained optimal power
  flow with renewables: A conic reformulation,'' \emph{IEEE Transactions on
  Power Systems}, vol.~33, pp. 1860--1867, 2018.

\bibitem{Chen2018}
Y.~Chen \emph{et~al.}, ``A distributionally robust optimization model for unit
  commitment based on kullback-leibler divergence,'' \emph{IEEE Transactions on
  Power Systems}, vol.~33, pp. 5147--5160, 9 2018.

\bibitem{Zhou2020}
A.~Zhou \emph{et~al.}, ``A linear programming approximation of distributionally
  robust chance-constrained dispatch with wasserstein distance,'' \emph{IEEE
  Transactions on Power Systems}, vol.~35, pp. 3366--3377, 9 2020.

\bibitem{KundurBook}
P.~Kundur, \emph{Power System Stability and Control}, 1st~ed.\hskip 1em plus
  0.5em minus 0.4em\relax McGraw-Hill Education, 1994.

\bibitem{Chavez2014}
H.~Chavez \emph{et~al.}, ``Governor rate-constrained {OPF} for primary
  frequency control adequacy,'' \emph{IEEE Transactions on Power Systems},
  vol.~29, pp. 1473--1480, 2014.

\bibitem{Sturt2012}
A.~Sturt and G.~Strbac, ``{Efficient stochastic scheduling for simulation of
  wind-integrated power systems},'' \emph{IEEE Transactions on Power Systems},
  vol.~27, no.~1, pp. 323--334, feb 2012.

\bibitem{Badesa2020}
L.~Badesa \emph{et~al.}, ``{Optimal Portfolio of Distinct Frequency Response
  Services in Low-Inertia Systems},'' \emph{IEEE Transactions on Power
  Systems}, vol.~35, no.~6, pp. 4459--4469, 2020.

\bibitem{boyd2004convex}
S.~Boyd and L.~Vandenberghe, \emph{Convex Optimization}.\hskip 1em plus 0.5em
  minus 0.4em\relax Cambridge university press, 2004.

\bibitem{truedata}
\BIBentryALTinterwordspacing
``{Electric Chargepoint Analysis 2017: Domestics},'' UK Department for
  Transport, Tech. Rep., Feb 2018. [Online]. Available:
  \url{https://www.gov.uk/government/statistics/electric-chargepoint-analysis-2017-domestics}
\BIBentrySTDinterwordspacing

\bibitem{Sturt_conf}
A.~Sturt and G.~Strbac, ``A times series model for the aggregate {GB} wind
  output circa 2030,'' in \emph{IET Conference on Renewable Power Generation
  (RPG 2011)}, 2011, pp. 1--6.

\bibitem{PFENNINGER20161251}
S.~Pfenninger and I.~Staffell, ``{Long-term patterns of European PV output
  using 30 years of validated hourly reanalysis and satellite data},''
  \emph{Energy}, vol. 114, pp. 1251--1265, 2016.

\end{thebibliography}

\begin{IEEEbiography}
    [{\includegraphics[width=1in,height=1.25in,clip,keepaspectratio]{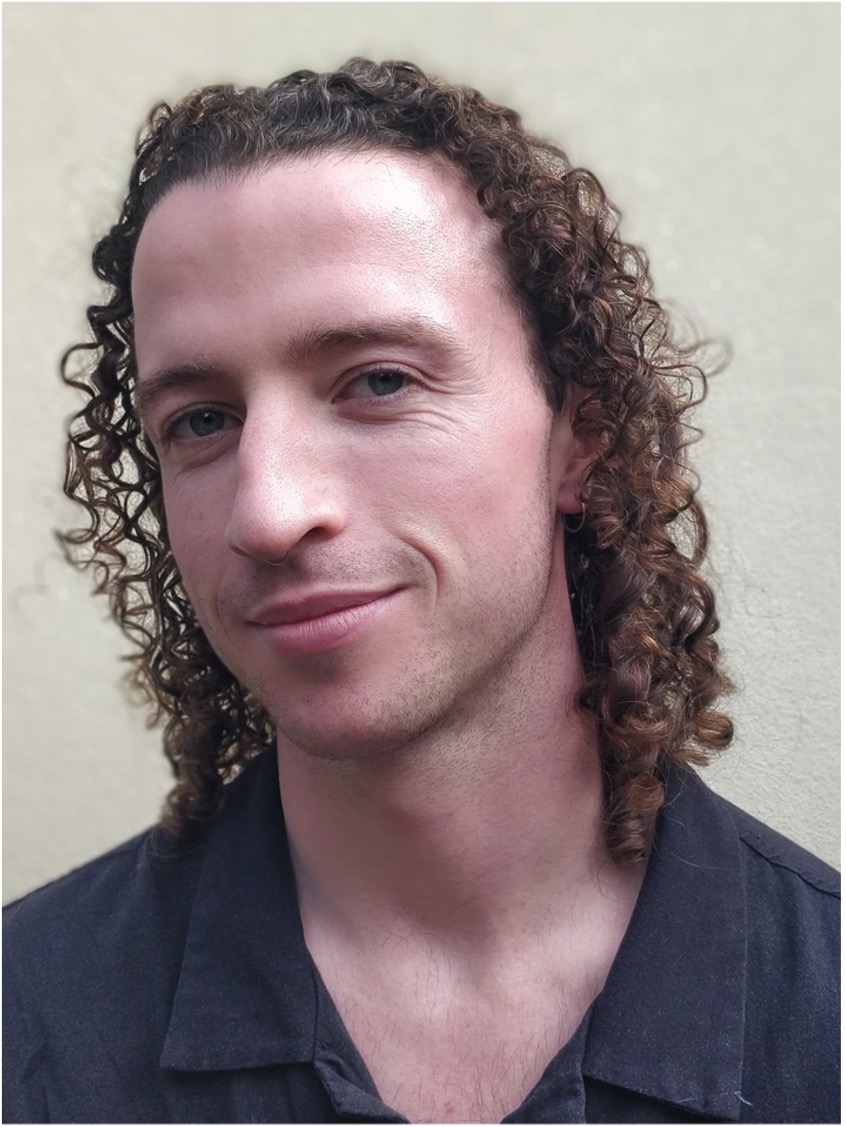}}]{Cormac O'Malley}
(S'18) received the MEng degree in Engineering Science from the University of Oxford, U.K, in 2018. He is currently pursuing a Ph.D. in Electrical Engineering at Imperial College London, U.K. His research interests lie in modelling and optimisation of low carbon power grid operation. 
\end{IEEEbiography}

\begin{IEEEbiography}
    [{\includegraphics[width=1in,height=1.25in,clip,keepaspectratio]{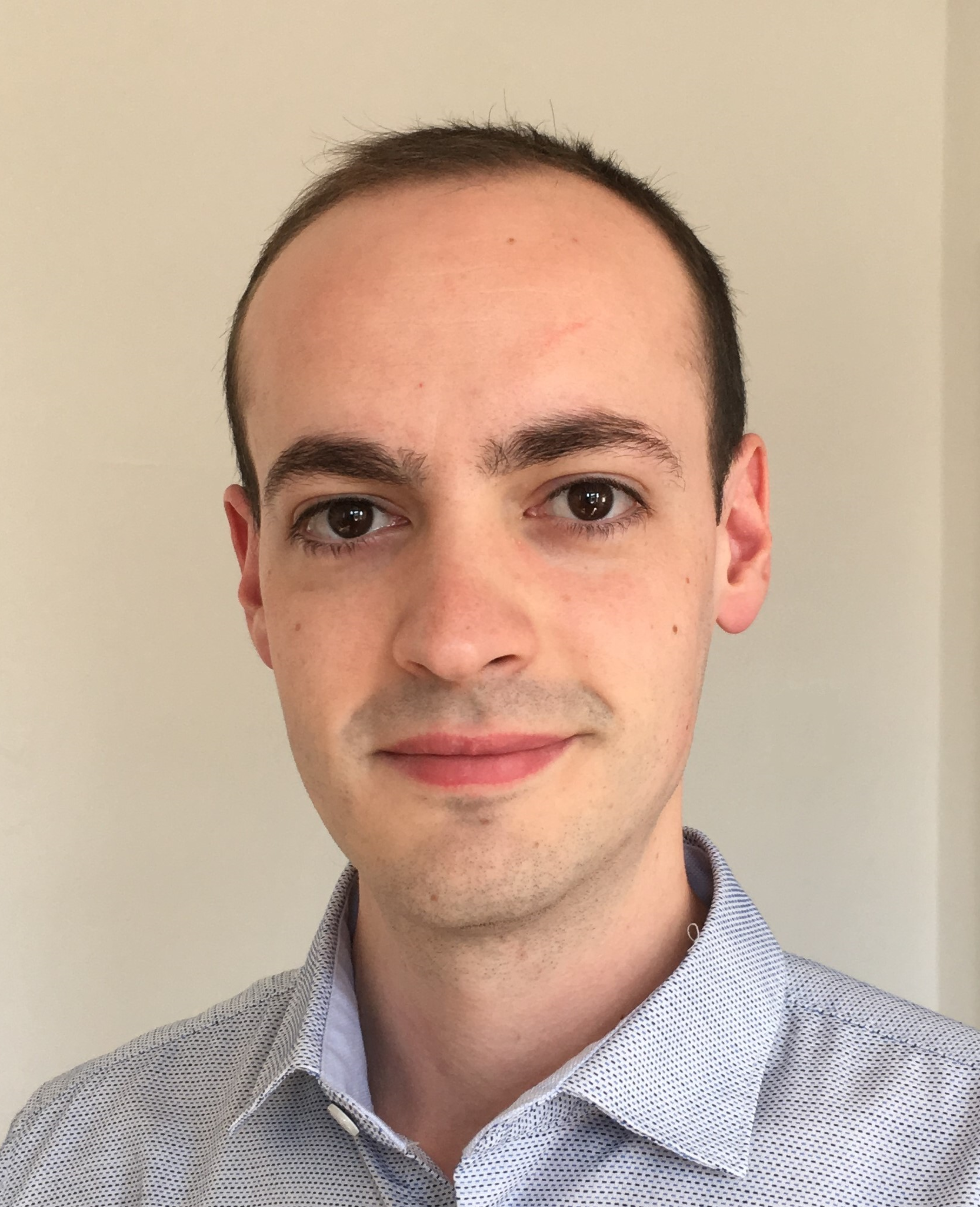}}]{Luis Badesa}
(S'14-M'20) received the Ph.D. degree in Electrical Engineering from Imperial College London, U.K., in 2020. He will start as Assistant Professor in Electrical Engineering at the Technical University of Madrid (UPM), Spain, and is currently a Research Associate at Imperial College London. His research focus is on modelling the operation and economics of low-inertia electricity grids, and market design for frequency-containment services.
\end{IEEEbiography}

\vspace{14cm}

\begin{IEEEbiography}
    [{\includegraphics[width=1in,height=1.25in,clip,keepaspectratio]{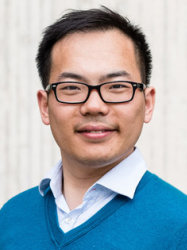}}]{Fei Teng}
(M'15) received the Ph.D. degree in Electrical Engineering from Imperial College London, U.K, in 2015. Currently he is a Lecturer in the Department of Electrical and Electronic Engineering, Imperial College London, U.K. His research focuses on scheduling and market design for low-inertia power systems, cyber-resilient energy system operation and control, and objective based data analytics for future energy systems.
\end{IEEEbiography}

\vspace{-14cm}

\begin{IEEEbiography}
    [{\includegraphics[width=1in,height=1.25in,clip,keepaspectratio]{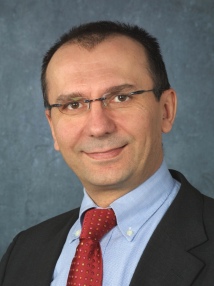}}]{Goran Strbac}
(M'95) is Professor of Electrical Energy Systems at Imperial College London, U.K. His current research is focused on optimisation of operation and investment of low-carbon energy systems, energy infrastructure reliability and future energy markets.
\end{IEEEbiography}

\end{document}